\documentclass[final]{jfm}

\usepackage{graphicx}
\usepackage{subcaption} 
\usepackage{overpic} 
\usepackage{newtxtext}
\usepackage{newtxmath}
\usepackage{natbib}
\usepackage{hyperref}
\usepackage{bm}
\usepackage{float}
\usepackage{mathtools}
\usepackage{caption}
\hypersetup{
    colorlinks = true,
    urlcolor   = blue,
    citecolor  = blue,
}

\usepackage{xcolor}
\newcommand{\RomanNumeralCaps}[1]
\nolinenumbers

\title{Supersonic jet dynamics from two-cavitation-bubble interactions: acceleration, tip fragmentation and penetration}

\author{Shuai Yan\aff{1,2},
	A-Man Zhang\aff{1,2}	\corresp{\email{zhangaman@hrbeu.edu.cn}},
	Tianyuan Zhang\aff{1,2},
	Pu Cui\aff{1,2},
	Rui Han\aff{3},
	\corresp{\email{hanrui@hrbeu.edu.cn}},
	\and Shuai Li\aff{1,2}	\corresp{\email{lishuai@hrbeu.edu.cn}}}

\affiliation{\aff{1}College of Shipbuilding Engineering, Harbin Engineering University, Harbin 150001, PR China
\aff{2}National Key Laboratory of Ship Structural Safety, Harbin Engineering University, Harbin 150001, PR China
\aff{3}Heilongjiang Provincial Key Laboratory of Nuclear Power System and Equipment, Harbin Engineering University, Harbin 150001, PR China
}

\begin{document}
\maketitle

\begin{abstract}
This study experimentally and numerically investigates the dynamics of a high-speed liquid jet generated from the interaction of two tandem cavitation bubbles, termed bubble 1 and bubble 2, depending on their generation sequence. Although the overall collapse pattern and jet orientation are well documented, the underlying mechanisms for supersonic jet acceleration, tip fragmentation, and subsequent penetration remain to be elucidated. In our experiments, two near-identical, highly-energized cavitation bubbles were generated using an underwater electric discharge method, and their transient interactions were captured using a high-speed camera. We identify three distinct jet regimes that emerge from the tip of bubble 2: conical, umbrella-shaped, and spraying jets, characterized by variations in the initial bubble-bubble distance (denoted as $\gamma$) and the initiation time difference (denoted as $\theta$). Our numerical simulations using both Volume of Fluid and Boundary Integral methods reproduce the experimental observations quite well and explain the mechanism of jet acceleration. We show that the transition between the regimes is governed by the spatiotemporal characteristics of the pressure wave induced by the collapse of bubble 1, which impacts the high-curvature tip of bubble 2. Specifically, a conical jet forms when the pressure wave impacts the bubble tip prior to its contraction, while an umbrella-shaped jet develops when this impact occurs after the contraction. The spraying jets result from the breakup of the bubble tip, exhibiting mist-like and needle-like morphologies with velocities ranging from 10 to over 1200 m/s. Remarkably, we observe that the penetration distance of spraying jets exceeds ten times the maximum bubble radius, making them ideal for long-range, controlled fluid delivery. Finally, phase diagrams for jet velocity and penetration distance in the $\gamma-\theta$ parameter space are established to provide a practical reference for biomedical applications, such as needle-free injection and micro-pumping.
\end{abstract}

\begin{keywords}
bubble dynamics, cavitation, jets
\end{keywords}

{\bf MSC Codes }  {\it(Optional)} Please enter your MSC Codes here

\section{Introduction}
\label{sec:headings}

Cavitation bubbles, with their unique transient dynamic characteristics such as shock waves and high-speed jets during collapse, play a pivotal role in the destructive phenomenon of hydraulic machinery erosion \citep{blake1987cavitation,brennen1995cavitation,philipp1998cavitation,zhao2025new} and beneficial applications like ultrasonic cleaning \citep{ohl2006surface}, ultrasonic emulsification \citep{ren2023interactions,li2024cavitation}, sonochemistry \citep{suslick1990sonochemistry,moholkar1999hydrodynamic}, etc. Liquid jets produced by bubbles near boundaries or under ultrasonic excitation are of significant scientific interest because their localized high-velocity flows can markedly enhance mass transport in liquids \citep{yang2023enhanced, li2024cavitation} and enable precise perforation of soft materials \citep{arita2013laser, wei2014laser, orthaber2014effect, george2018minireview, wang2022experimental, cattaneo2025cyclic}. Recent studies have revealed a singular jet with velocity exceeding 1000 m/s generated by a single bubble near a rigid boundary \citep{lechner2019fast, lechner2020jet, reuter2021supersonic}, nearly ten times that of conventional bubble jets. Generating such extremely high velocities requires a very small initial bubble-wall standoff distance. \citet{brujan2001dynamics} also found that the jet velocity can reach 960m/s for a laser-induced cavitation bubble near an elastic boundary. However, due to the stringent boundary conditions and the risk of shock waves harming neighboring targets or tissues \citep{ohl2003detachment, dijkink2007controlled}, single-bubble systems are less controllable for practical applications like needle-free injection.

A more controlled, slender, and high-speed piercing jet can be achieved through the interaction between two tandem bubbles with temporal offset \citep{han2015dynamics,liang2021interaction,fan2024amplification}. The dynamics of the tandem bubbles and associated piercing jets have been investigated across millimeter scale to centimeter scale using laser pulses \citep{han2015dynamics, tomita2017pulsed, robles2020soft, fan2024amplification} and electrical discharge methods \citep{fong2009interactions, luo2021stratification, liang2021interaction, liang2022experimental}. To ensure clarity throughout this paper, the temporally preceding bubble will be referred to as bubble 1, while the subsequent bubble will be designated as bubble 2. The overall two-bubble interaction, collapse pattern, and jet orientation have been well documented in the published literature. The foundational study by \citet{fong2009interactions} first delineated the "catapult" effect through experiments, uncovering a two-phase hydrodynamic process: (i) bubble 2 is initially drawn toward bubble 1, and (ii) it is then propelled away, leading to a high-speed jet that pierces through bubble 2. In this process, bubble 1 acts as a "free surface" for bubble 2, thereby inducing a reverse jet that is directed away from this "free surface." \citet{han2015dynamics} provided a more systematic analysis by examining the effects of the initial time difference $\Delta t$ and the initial bubble distance $d$ on tandem bubble dynamics. They used synchronized laser-induced bubble experiments and numerical simulations, focusing on the velocity changes of the jet from the second bubble. This type of directed jet has been successfully applied for the precise penetration of soft biomaterials \citep{robles2020soft}.

The two-bubble system produces a jet with distinct advantages over a single-bubble jet, including a more localized impact area and greater penetration depth. It also has the potential to reduce thermal damage and lessen the impact of shock wave pressures on surrounding targets, due to the jet's ability to penetrate a relatively long distance. \citet{sankin2010pulsating} and \citet{yuan2015cell} showed that jets from bubble 1 can achieve localized cell membrane perforation when a single bubble cannot. This reinforces a better penetration of directed jets from bubble pairs over single bubbles. Recently, \citet{fan2024amplification} reported the formation of supersonic microjets with velocities exceeding 1000 m/s through a tandem bubble system. They revealed that the collapse of bubble 1 induces radial contraction and subsequent pinch-off of the elongated surface of bubble 2, a process referred to as neck breakup. The intense focusing flow creates a high-pressure stagnation point that further accelerates the fluid into supersonic microjets within bubble 2. Nevertheless, we need a more in-depth understanding of jet acceleration, jet tip fragmentation, and subsequent penetration of the jet, all of which are crucial for expanding their potential in needle-free biomedical applications.
\begin{figure}
\centerline{\includegraphics[width=0.6\textwidth]{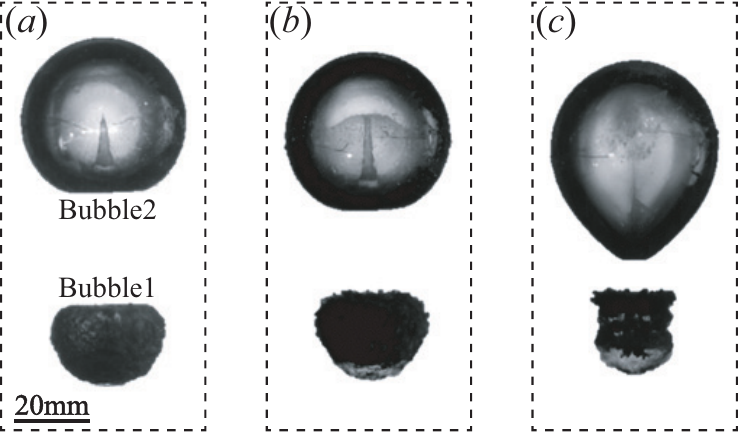}}
  \caption{Three regimes of piercing jets formed under the interaction of two tandem bubbles. The dynamics of the two tandem bubbles are governed primarily by their initial spatial offset $d$ and the temporal delay $\Delta t$. (\textit{a}) Conical jet, with $d$ = 45.1 mm and $\Delta t$ = 2.40 ms; (\textit{b}) Umbrella-shaped jet, with $d$ = 37.5 mm and $\Delta t$ = 3.00 ms; and (\textit{c}) Spraying jet,  with $d$ = 36.7 mm and $\Delta t$ = 2.80 ms. The scale bar in the figure represents a length of 20 mm.}
\label{fig:1}
\end{figure}

In this study, we conduct hundreds of experiments on two-bubble interactions using the underwater electric discharge method \citep{Cui2020, han2022interaction, zhang2025free}. We identify three distinct piercing jet regimes that emerge from the bottom of bubble 2, as illustrated in Figure \ref{fig:1}(\textit{a})-(\textit{c}), which correspond to conical, umbrella-shaped, and spraying jets. We reveal the dependencies of jet features on the initial bubble-bubble distance $d$ and the initiation time difference $\Delta t$ between the two bubbles. Additionally, our numerical simulations, employing both the Volume of Fluid and Boundary Integral methods, reproduce the experimental observations well. By combining experiments and numerical simulations, this study aims to (i) elucidate the underlying mechanisms of piercing jet acceleration, (ii) reveal the jet regimes across a large parameter space, (iii) quantify their penetration capabilities in different regimes, and (iv) establish an optimal control strategy for controlling jet morphology, velocity, and penetration distance in a two-bubble system.

The structure of this paper is organized as follows. First, the experimental and numerical methodologies are detailed in \S{2}. In \S{3}, we present the formation and evolution of the three regimes of piercing jets, along with a comparison between experimental observations and numerical results. In \S{4}, we examine how the collapse of bubble 1 governs the initiation of conical and umbrella-shaped jets of bubble 2. In \S{5}, we illustrate the formation and instability of spraying jets. In \S{6}, a simplified liquid-bullet model is employed to investigate the evolution of different types of piercing jets after they penetrate the bubble surface. Meanwhile, the phase diagrams for jet velocity and penetration distance in the governing parameter space are established. We provide a quantitative discussion of how the governing parameters influence different regimes of piercing jets. Finally, conclusions derived from this study are summarized in \S{7}.

\section{Methodology}\label{2}

\begin{figure}
	\begin{subfigure}[c]{0.56\textwidth} 
		\centering
		\begin{overpic}[width=\textwidth]{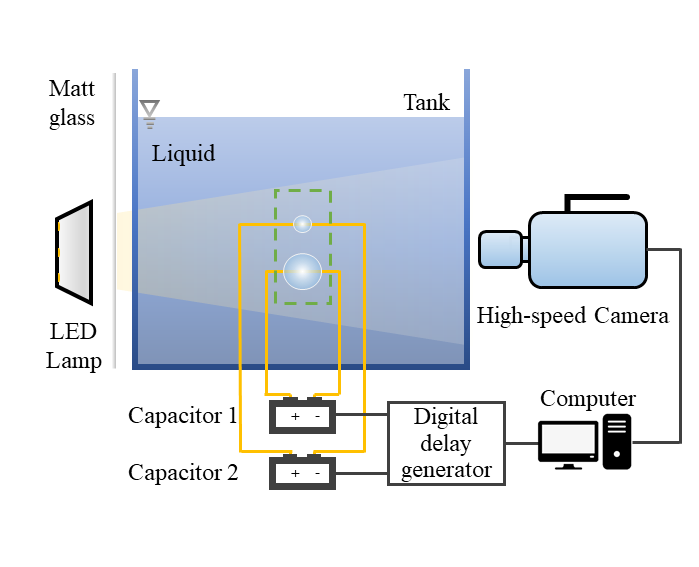} 
			\put(5,75){\large \textrm{(\it a\rm)}} 
		\end{overpic}
		\label{fig:sub1} 
	\end{subfigure}
	\begin{subfigure}[c]{0.45\textwidth} 
		\vspace{-0.8cm} 
		\begin{overpic}[width=\textwidth]{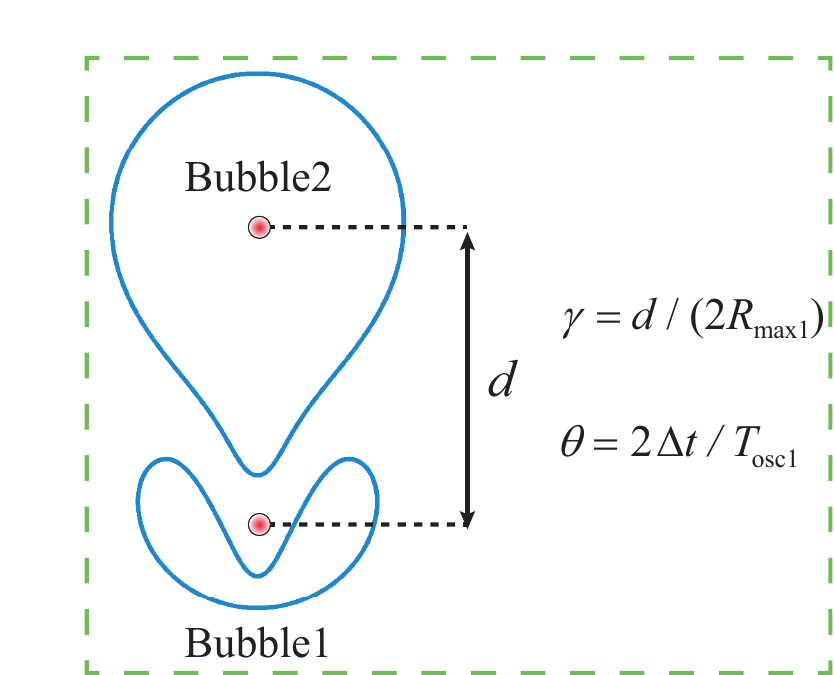} 
			\put(5,77.5){\large \textrm{(\it b\rm)}} 
		\end{overpic}
		\label{fig:sub2}
	\end{subfigure}
	\caption{($a$) Experimental set-up for bubble pair dynamics. Centimetre-scale bubbles are generated by an underwater electric discharge method. A computer and a digital delay generator are used to control the discharge process of the two capacitors with a preset time delay. ($b$) Illustration of parameters governing dynamics of two tandem bubbles, with \textit{d} denoting the relative initiation bubble distance, and \textit{$R_{\rm max}$} representing the maximum radius of the two same-sized bubbles. Bubble 2 is generated after a time delay \textit{$\Delta t$} following the initiation of bubble 1. $T_{\rm osc1}$ represents the period of bubble 1 oscillation, defined as the time from bubble generation to the first collapse.}
	\label{fig:two_images}
\end{figure}

\subsection {Experimental set-up}
The experiments of two-bubble interactions are performed in a water tank of 600 mm × 600 mm × 600 mm under atmospheric pressure (\textasciitilde 101 kPa) and at room temperature (\textasciitilde $20^\circ \rm C$). An underwater electric discharge method \citep{ Cui2020, han2022interaction, zhang2025free, pei2025influence} is adopted to generate cavitation bubbles as sketched in Figure \ref{fig:two_images}(\textit{a}). The system utilizes two parallel 2200 \textmu F capacitors whose positive and negative electrodes are independently wired. The wire radius used is approximately 0.1 mm, which corresponds to less than 0.05\% of the maximum bubble radius. Its influence on bubble pulsation and jetting behavior is negligible \citep{Li2020}. Two localized high-resistance points are intentionally introduced at the midpoint of the wires to facilitate bubble generation through concentrated Joule heating during the discharge process \citep{Cui2020}. In all experiments, identical discharge voltages are applied to both capacitors to ensure consistent bubble sizes between bubble 1 and bubble 2. Experiments are conducted at a constant discharge voltage of 500 V. Through repeated experiments, the average maximum bubble radius $R_{\rm max}$ is determined to be 23 mm, with the corresponding first oscillation period measuring approximately 4.5 ms. Here, $T_{\rm osc}$ is defined as the time from bubble generation to the first collapse, identified via high-speed imaging.
The bubble size distribution in repeated experiments exhibits a maximum radius variation of less than 5\%, well within the 15\% threshold suggested by \citet{fong2009interactions} and \citet{han2015dynamics} as acceptable for the claim of similarly sized bubbles. The initial spacing between the two bubbles can be easily controlled by adjusting the relative distance between the two localized high-resistance points. The timings of the two bubbles are precisely managed using a digital delay generator (DG645, SRS Stanford Research Systems), starting with the discharge of capacitor 1 followed by capacitor 2. Accounting for the temporal uncertainty of the discharge circuit, the predetermined delay interval between the two bubbles can be adjusted to within ±2 \textmu s precision, which is negligible compared to the bubble oscillation period. The cavitation bubbles are initially positioned at a water depth of approximately 300 mm ($\textgreater 13R_{\rm max}$), where boundary effects play a minor role in the dominant bubble-bubble interactions \citep{supponen2016scaling}.

A high-speed camera (Phantom V2012), working at 43000-110000 frames per second with an exposure time of 1-2 \textmu s, is employed to capture the transient bubble behaviors. A macro lens with 100 mm focal length is employed to resolve the bubble dynamics in detail, while another lens with 50 mm focal length captures the propagation of the piercing jet in the liquid. A continuous light source (600 W) filtered through matt glass provides illumination from the back. Time zero is defined as the instant of bubble 1 initiation, coinciding with the last shadowgraph frame captured before discharge spark emergence. Consequently, the temporal offset between the defined time zero and actual bubble inception is bounded by the maximum frame interval of 9-23 \textmu s. The temporal resolution (9-23 \textmu s) is within 1\% of the first period of bubble oscillation, which becomes negligible relative to the oscillation timescale about 4 ms. 
Length measurements are subject to the spatial uncertainty of one image pixel (0.1-0.2 mm), corresponding to approximately 1\% of the maximum bubble radius.

\subsection {Numerical methods}
To investigate the underlying mechanisms governing the interaction of two tandem bubbles, we employ two sets of complementary numerical models based on the Boundary Integral Method (BIM) and Volume of Fluid (VoF) for numerical simulations with different purposes. First, the BIM offers computational flexibility to decouple complex bubble interactions through numerical manipulations (such as artificial adjustment of bubble morphology or position), thereby isolating dominant factors governing piercing jet formation \citep{fong2009interactions,dadvand2012boundary,Peters2013,han2022interaction}.
However, after the jet penetration, the vortex ring model needs to be employed to continue the simulation \citep{wang1996nonlinear,han2022interaction,li2024cavitation}, where topological complexity introduces significant computational challenges. 
Therefore, we utilize a complementary VoF model, employing grid-based simulations to represent the complete bubble evolution while capturing flow features inaccessible to experiments.

In the first model, we adopt a well-verified boundary integral code based on potential flow theory \citep{han2022interaction,li2024cavitation}. The flow surrounding the bubbles satisfies the Laplace equation, while the gas pressure inside each bubble is assumed spatially uniform and governed by the adiabatic equation of state. The simulation initiates both bubbles simultaneously while initially positioning bubble 2 at infinity with frozen potential (no velocity potential updating) to eliminate its hydrodynamic influence on bubble 1's oscillation. Upon reaching the predetermined delay time, bubble 2 is relocated to the specified position to activate bubble interactions. We utilize 200 surface elements per bubble, a configuration that optimally balances numerical accuracy with computational efficiency \citep{Li2020, Tong2022, yan2025numerical}.

In the second model, we employ VoF to track the gas-water interface and utilize the Finite Volume Method to solve the Navier-Stokes equations, capturing the complete evolution process of two tandem bubbles, including the jet penetration and breakup. An improved and well-validated solver cavBubbleFoam \citep{koch2016numerical, zeng2018wall, reese2022microscopic}, derived from the compressibleInterFoam solver in the open-source platform OpenFOAM \citep{weller1998tensorial}, is utilized to perform the simulations. In the VoF framework, the two fluids are treated as immiscible. The liquid phase volume fraction ($ \alpha $) and the gas phase volume fraction ($1-\alpha$) are governed by a transport equation
\begin{equation}
\frac{\rm \partial \it \alpha}{\partial t} + \nabla \cdot (\alpha \textbf{\textit{u}}) + \nabla \cdot \left( \alpha (1 - \alpha) \bm{\mathit{U_{r}}} \right) = \alpha (1 - \alpha) \left( \frac{\psi_g}{\rho_g} - \frac{\psi_l}{\rho_l} \right) \frac{\rm D \it p}{\rm D \it t} + \alpha \nabla \cdot \textbf{\textit{u}},
\end{equation}
where \textbf{\textit{u}} denotes the velocity field, $t$ is time, $\rho$ represents density and $\psi=\rm D\it \rho/\rm D \it p$, with subscripts $l$ and $g$ distinguishing the liquid and gas phases. The $\bm{\mathit{U_{r}}}$ is the relative velocity between the two phases and the term $\nabla \cdot \left( \alpha (1 - \alpha) \bm{\mathit{U_{r}}} \right)$ is used to guarantee a sharp interface \citep{rusche2002computational,deshpande2012evaluating}.
At each time step, both the liquid and gas phases satisfy the Navier–Stokes equations:
\begin{equation}
\frac{\partial \rho}{\partial t} + \nabla \cdot (\rho \textbf{\textit{u}}) = 0,
\end{equation}
\begin{equation}
\frac{\partial \rho \textbf{\textit{u}} }{\partial t} +\nabla \cdot (\rho \textbf{\textit{uu}})=-\nabla p+\nabla \cdot \textbf{\textit{S}}+\textbf{\textit{f}}_{\delta }.
\label{eq.2}
\end{equation}
The viscous stress tensor \( \textbf{\textit{S}} \) is defined by \( \textbf{\textit{S}}=\mu (\nabla \textbf{\textit{u}}+\nabla \textbf{\textit{u}}^\mathrm{T}-2(\nabla \cdot \textbf{\textit{u}})\textbf{\textit{I}}/3) \), where \( \mu \) denotes the dynamic viscosity and \( \textbf{\textit{I}} \) the identity tensor. The term \( \textbf{\textit{f}}_{\delta } \) accounts for surface tension contributions. Numerical implementation follows the methodology detailed in \citet{zeng2018jetting} and \citet{reese2022microscopic}.

The governing equations are thermodynamically closed using the Tait equation of state for both the gas inside the bubble and the surrounding liquid

\begin{equation}
p = \left( p_0 + B \right) \left( \frac{\rho}{\rho_0} \right)^\kappa - B,
\label{2.4}
\end{equation}
where reference pressure \( p_0 = 101325 \, \text{Pa} \), Tait pressure \( B = 304.6 \, \text{MPa} \), reference density \( \rho_0 = 998 \, \text{kg} \, \text{m}^{-3} \) and adiabatic coefficient \( \kappa = 7.15 \) for the liquid (water). 
The gas inside the bubble is modeled as an ideal gas using the adiabatic equation of state (a special case of the Tait equation with \( B = 0 \)), with \( p_0 = 101325 \, \text{Pa} \), \( \rho_0 = 1.29 \, \text{kg} \, \text{m}^{-3} \) and \( \kappa = 1.25 \) \citep{fong2009interactions,han2022interaction}.

The mesh size in the computational domain encompassing both the bubbles and jet is set to 1/460 of the maximum bubble radius to ensure mesh convergence \citep{zeng2018wall,reese2022microscopic}. At initiation, the two tandem bubbles are modeled as spherical with radii ranging from 2.5 to 6.7 mm and internal pressures between 2 and 28 MPa, stationary in water. The bubble size and period predicted by the Keller–Miksis equation \citep{keller1980bubble} under these conditions show slight deviation from experimental measurements, indicating that the results are robust to the choice of initial parameters within this range. Phase-change effects are taken into account as the equilibrium radius of bubble 1 is reduced by 65\% at its maximum \citep{koch2016numerical,liang2022comprehensive,fan2024amplification}. The numerical calculation domain of VoF is large enough (200$R_{\rm max}$) to mitigate the influence of reflected pressure waves on the core computational domain.

\subsection {Non-dimensionalization}
The system is non-dimensionalized using three fundamental quantities: the maximum radius of the two same-sized bubbles $R_{\rm max}$, the hydrostatic pressure $P_{\rm \infty}$, and the liquid density $\rho$. To elucidate the dominant factors governing bubble dynamics, we first examine the relevant dimensionless numbers. At the bubble scale, the characteristic Reynolds, Weber, and Froude numbers are respectively defined as:
\begin{align}
{\small Re = \frac{\rho U R_{\rm max}}{\mu} \sim O(10^5), \quad We = \frac{\rho U^2 R_{\rm max}}{\sigma} \sim O(10^4), \quad Fr = \frac{U}{\sqrt{gR_{\rm max}}} \sim O(10),}
\end{align}
where $U=\sqrt{P_{\rm \infty}/\rho}\approx 10 \rm ~m/s$ is the characteristic velocity, the characteristic length of the bubble $R_{\rm max} \approx 23 ~\rm mm$, the water viscosity $\mu = 0.001 \rm ~Pa\cdot s$ and $\sigma = 0.073 \rm ~N/m$. This indicates that both viscous, capillary and gravity effects can be negligible. Therefore, these effects are excluded from our BIM calculations in \S{4}. The dynamics of the two tandem bubbles are governed primarily by their initial spatial offset $d$ and the temporal delay $\Delta t$. In our non-dimensional system, these parameters are presented as $\gamma$ and $\theta$,  
\begin{equation}
	\gamma = \frac{d}{2R_{\text{max1}}},~~~~~~ \theta = \frac{\Delta t}{R_{\rm max1} \sqrt{\rho / P_{\rm \infty}}},
\end{equation}
as shown in Figure \ref{fig:two_images}(\textit{b}).
For experiments presented in this study, we adopt the maximum width of bubble 1 as $R_{\rm max1}$ because of its spherical pulsation. This approximation is numerically validated to be reasonable, with the maximum width of bubble 1 deviating between 0.1\%-4\% compared to its free expansion value across the investigated parameter space ($0.7 < \theta < 1.7$ and $0.6< \gamma < 1$).
The $ 2R_{\rm max1} \sqrt{\rho / P_{\rm \infty}}$ values for all experimental data matches the first oscillation period of bubble 1 with maximum radius $R_{\rm max1}$ in a free-field under ambient pressure $ P_{\rm \infty}$, with deviations below 3\%. Therefore, this non-dimensionalization of the initiation time difference aligns with the approach by \citet{fong2009interactions} and \citet{han2015dynamics}, where the first bubble oscillation period serves as the characteristic timescale.

\section{Three Regimes of Jet Morphology}\label{3}
We first present an overview of the piercing jets observed in the experiments, supported by numerical simulations performed using OpenFOAM. The characteristics of different piercing jet regimes are analysed, followed by a discussion on their penetration performance in the liquid. 
\subsection {Conical jet}

\begin{figure}
	\vspace*{0mm}
	\centerline{\includegraphics[trim={15pt 0 10pt 75pt}, 
		clip, width=0.95\textwidth]{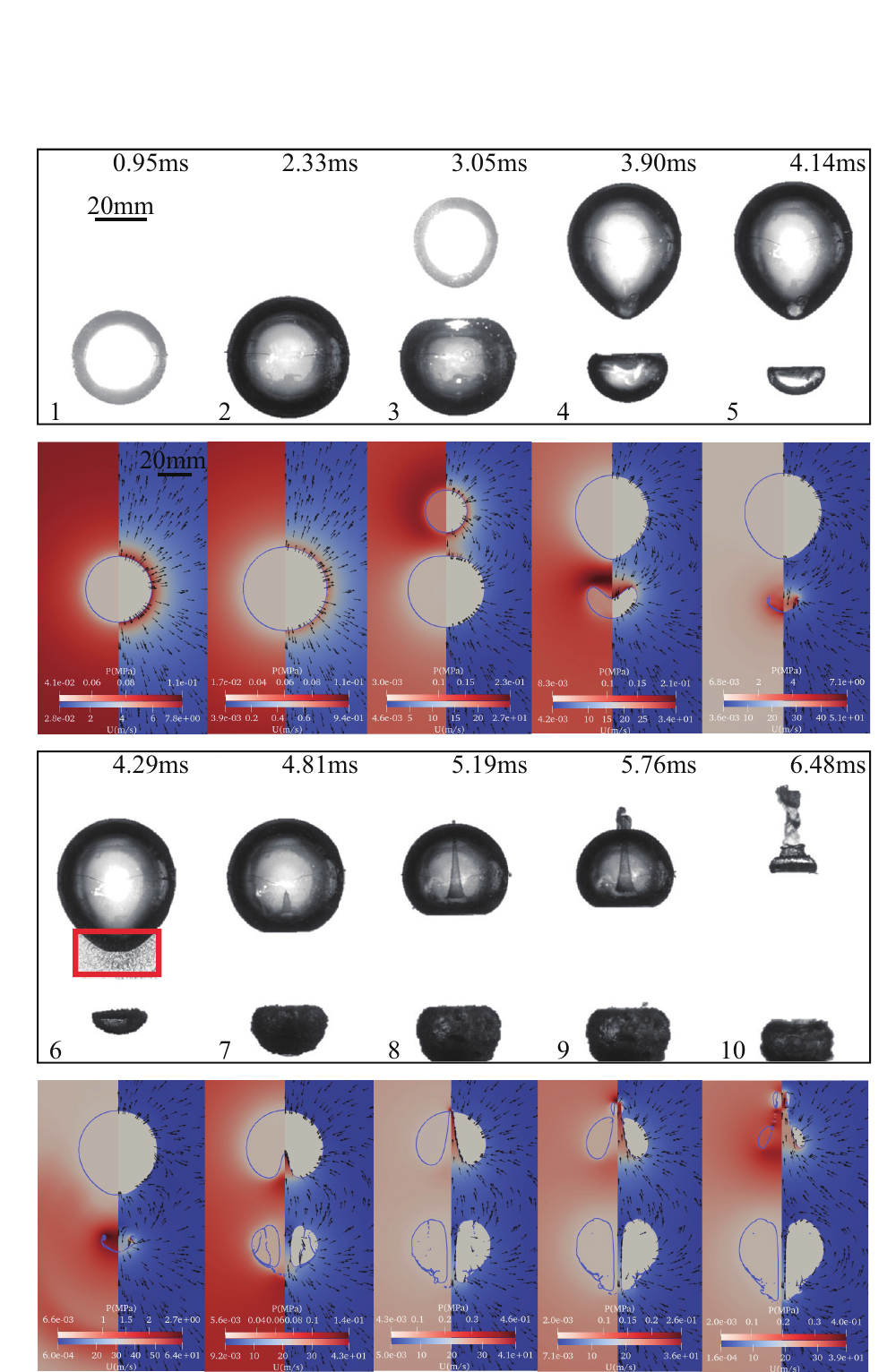}}
	\caption{Dynamics of conical jet from anti-phase bubble pair: comparison between the experiments (odd rows) and numerical simulations (even rows). The red box in frame 6 indicates a group of cavitation bubbles induced by the rarefaction wave, which is reflected off the interface of bubble 2 due to the collapsing shock wave from bubble 1. The dimensionless time corresponding to the numerical results are 0.42, 1.04, 1.36, 1.73, 1.84, 1.87, 2.13, 2.40, 2.56, and 2.76, respectively. The time scale $ R_{\rm max} \sqrt{\rho / P_{\rm \infty}}$ is 2.25ms. The initial parameters are $\gamma=0.98$, $\theta=1.10$.}
	\label{fig:3}
\end{figure}

Figure \ref{fig:3} illustrates the formation of a conical jet generated by two tandem bubbles with $\gamma=0.98$ and $\theta=1.10$, comparing the results from numerical simulations with experimental observations. Driven by its initial high pressure, bubble 1 expands spherically (frame 1) and imparts significant inertia to the surrounding fluid. This inertial causes the bubble to grow and exceed its equilibrium radius, depressing both its internal pressure and the pressure in the adjacent field to a level significantly below the local hydrostatic pressure when the bubble reaches its maximum radius at $t$ = 2.33 ms (frame 2). Subsequently, bubble 2 forms and expands in a non-spherical shape because of the asymmetric pressure distribution in the flow field relative to the center of bubble 2. The lower surface of bubble 2 is drawn toward bubble 1, forming a high-curvature tip (frames 3-5). The curvature at this tip is 5.87 times greater than that of the upper surface of bubble 2. Before the lower tip of bubble 2 fully contracts, the collapse of bubble 1 emits a shock wave that impacts the lower surface of bubble 2 (frame 6). This interaction generates a reflected rarefaction wave, which induces a cluster of micro-sized cavitation bubbles near the bottom of bubble 2, as highlighted by the red box in frame 6. The combined effect of the shock wave from the collapse of bubble 1 and the localized high-pressure region beneath bubble 2 further accelerates its lower surface, ultimately creating a distinct conical liquid jet (frame 7). As bubble 1 re-expands, its internal pressure drops rapidly (frame 7), consequently reducing the pressure in the surrounding flow field. However, the asymmetric collapse of bubble 2 triggers convergence of the surrounding fluid toward the initial contraction region, increasing pressure at the jet base. This elevated pressure further accelerates fluid motion and bubble contraction, establishing a positive feedback loop. The resulting momentum focusing effect \citep{lauterborn1982cavitation, philipp1998cavitation, koukouvinis2016numerical} maintains the high-pressure zone, thereby sustaining jet development. As the conical jet impacts the opposite surface of bubble 2, its velocity decreases very fast (frames 8–9). Meanwhile, a sharp protrusion forms at the top of bubble 2, travels a considerable distance, and fragments into smaller pieces gradually (frame 10).

Notably, for cases where the jet tip can be clearly tracked across several consecutive frames, the velocity of this conical jet passing through the center of bubble 2 in this experiment is approximately 32.3 m/s, measured using high-speed imaging by tracking the position of the jet across four consecutive frames. The error in space is the length of a pixel, corresponding to approximately 0.2 mm. The error in time due to camera jitter is on the scale of nanoseconds and thus negligible. Therefore, the error is calculated as ±0.4 mm/0.069 ms = ±5.8 m/s. To account for the refractive effect of the gas-liquid interface on displacement measurements, the distance from the jet tip to the bubble center was multiplied by a correction factor of 1.33 \citep{philipp1998cavitation}, yielding a corrected jet velocity of approximately 43.1 ± 7.7 m/s. This value aligns closely with the numerical simulation, which predicts an instantaneous jet velocity of 42 m/s at the same position. Despite the remarkable agreement illustrated in Figure \ref{fig:3}, the primary mechanism driving the formation of the conical jet, whether it is the high curvature of the jet tip or the collapse-induced high pressure from bubble 1, remains unanswered and will be further analyzed in \S{4.1}.

\subsection {Umbrella-shaped jet}

\begin{figure}
\vspace*{0mm}
\centerline{\includegraphics[trim={25pt 0 25pt 130pt}, 
	clip, width=0.95\textwidth]{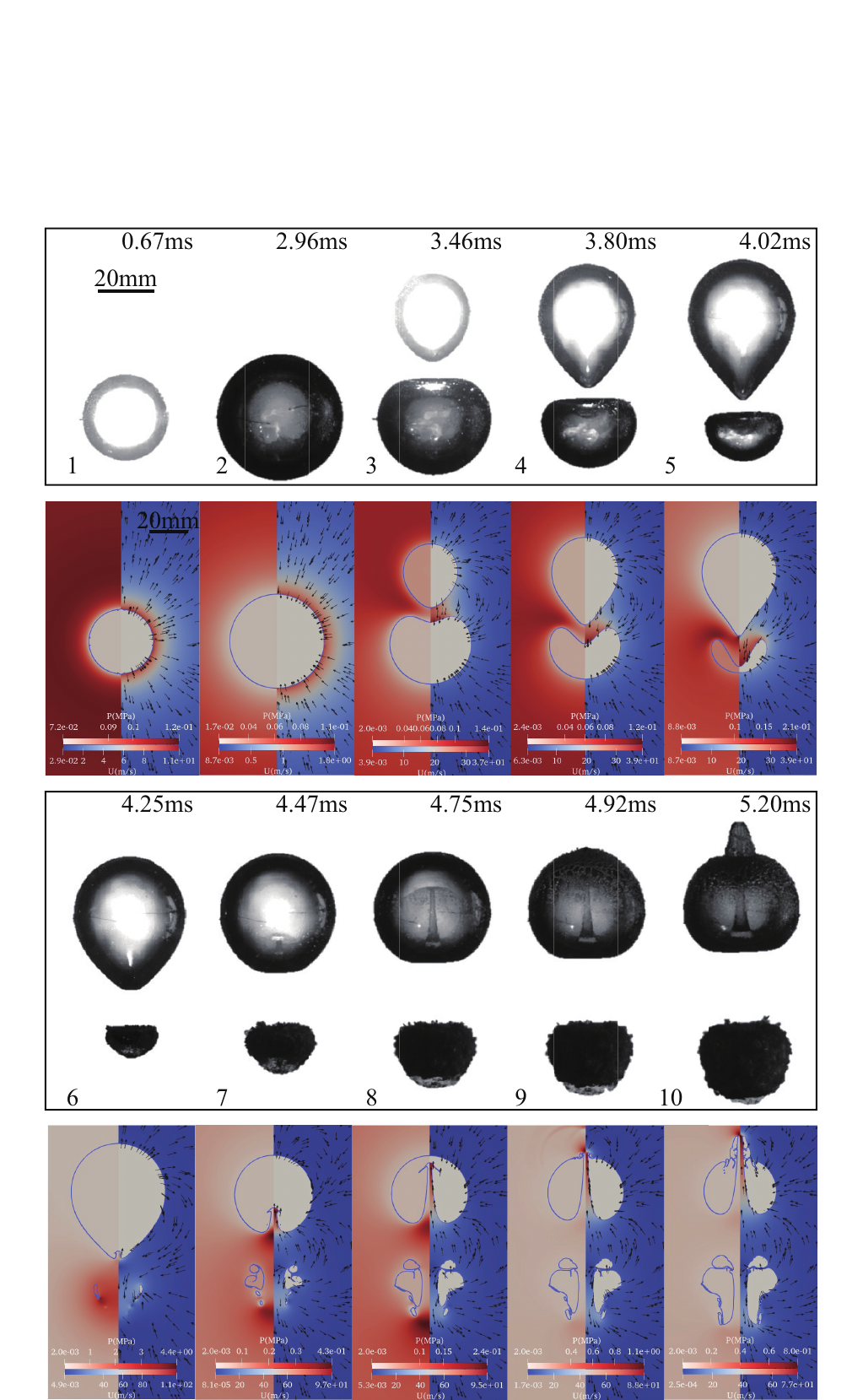}}
  \caption{Dynamics of umbrella-shaped jet from anti-phase bubble pair: comparison between the experiment (odd rows) and numerical simulation (even rows).  The dimensionless time corresponding to the numerical results are 0.29, 1.31, 1.51, 1.64, 1.78, 1.89, 1.98, 2.09, 2.13, and 2.22, respectively. The time scale $ R_{\rm max} \sqrt{\rho / P_{\rm \infty}}$ is 2.25ms. The initial parameters are $\gamma=0.82$, $\theta=1.38$.}
\label{fig:4}
\end{figure}

Decreasing the distance between the two tandem bubbles enhances their interaction, as shown in Figure \ref{fig:4}. This figure compares the experimental and numerical results for the umbrella-shaped jet with $\gamma=0.82$ and $\theta=1.38$. Similar to the initial evolution of the conical jet, the lower surface of bubble 2 elongates under the influence of bubble 1 (frames 1-4), forming a tip with a curvature that is 7.33 times that of the upper surface, which represents a notable increase compared to the case shown in Figure \ref{fig:3}. Driven by this increased curvature, the tip contracts during the collapse of bubble 1 (frames 5–6). As shown in the numerical results corresponding to frame 6, the umbrella-shaped jet tip forms before the final collapse of bubble 1. Following this, the toroidal collapse of bubble 1 generates multiple pressure shock waves \citep{ohl,Gonzalez2}. Given that the duration of these shock waves is substantially shorter than the characteristic time scale of jet evolution, they cumulatively accelerate the upstream fluid of the umbrella-shaped jet, causing it to widen or even form a second umbrella-shaped structure, which will be analysed in \S{4.2}. As shown in frames 7 and 8, the jet tip of bubble 2 takes on a clear umbrella-shaped morphology in the experiment, characterized by a flattened leading edge and a fragmented liquid film surrounding it. While similar jet morphologies have been documented in other systems \citep{shinjo2010simulation,karri2012jets,supponen2015inner,koukouvinis2016simulation,zhang2025free}, the underlying mechanism behind this umbrella-shaped jet and its dynamics remain poorly understood. A more detailed and in-depth analysis of these phenomena will be presented in \S{4.2}.

In contrast to the conical jet, the umbrella-shaped jet achieves a significantly higher peak velocity, reaching approximately 92.3 m/s in numerical simulations and 85.6 ± 7.7 m/s in experiments. Remarkably, the jet maintains stable propagation after penetrating the upper surface of bubble 2, traveling a distance of approximately 4.5 times the maximum radius of the bubble. The unique capability of high-velocity liquid jets to achieve substantial penetration distance enables the tandem bubble interaction to exhibit strong potential for needle-free injection and micro-pumping applications \citep{ohl2006sonoporation,robles2020soft,sankin2010pulsating}. We will further explore the dependence of the jet's penetration distance across a wide parameter space in \S{6}.

\begin{figure}
	\vspace*{0mm}
	\centerline{\includegraphics[trim={25pt 0 25pt 120pt}, 
		clip, width=0.85\textwidth]{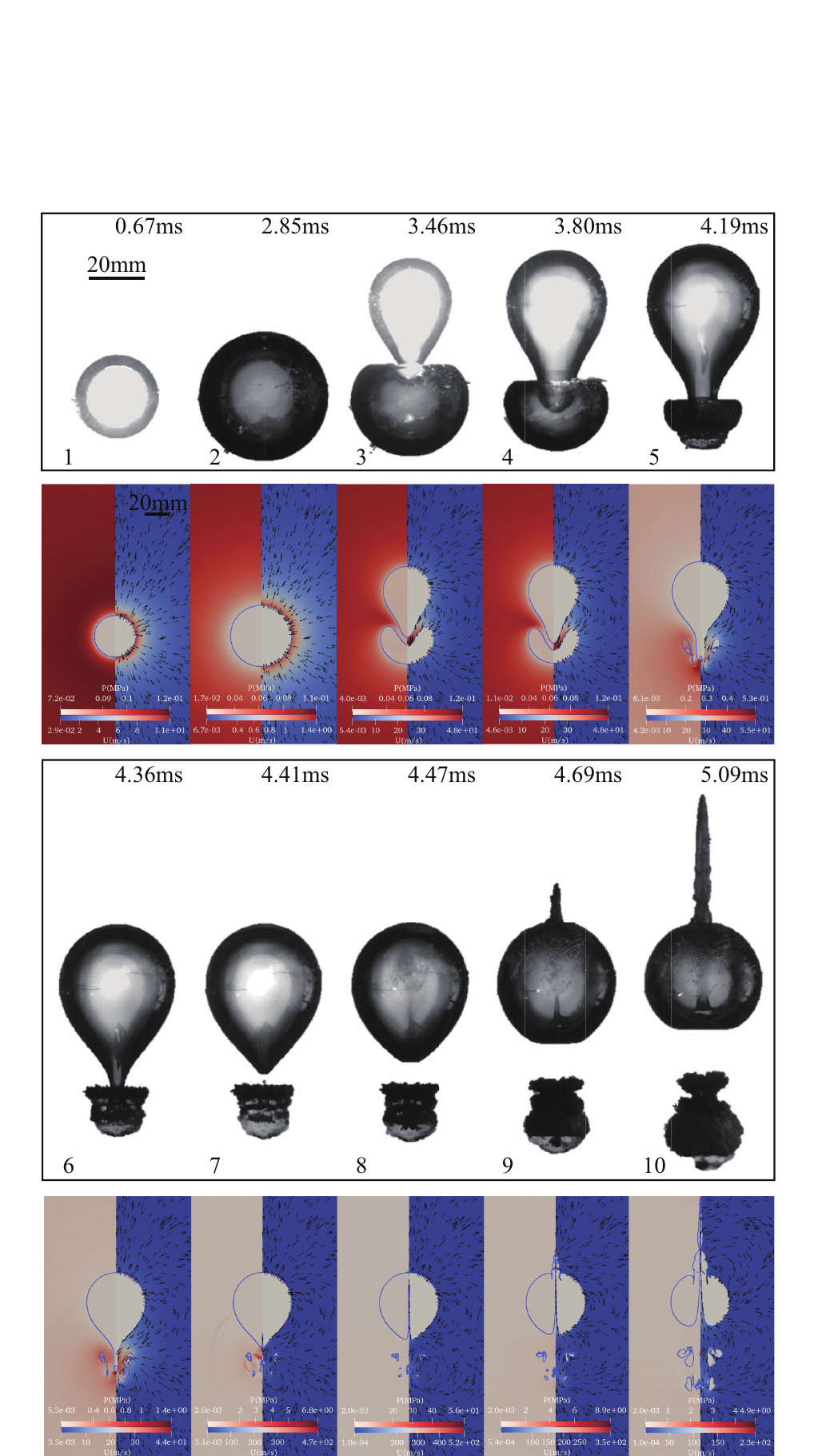}}
	\caption{Dynamics of spraying jet from anti-phase bubble pair: comparison between the experiment (odd rows) and numerical simulation (even rows).  The dimensionless time corresponding to the numerical results are 0.28, 1.21, 1.49, 1.53, 1.72, 1.77, 1.79, 1.83, 1.87, and 2.00, respectively. The time scale $ R_{\rm max} \sqrt{\rho / P_{\rm \infty}}$ is 2.35ms. The initial parameters are $\gamma=0.76$, $\theta=1.23$.}
	\label{fig:5}
\end{figure}
\subsection {Spraying jet}
Further enhancing the interaction between the two tandem bubbles leads to the transition of the jet morphology from an umbrella-shaped morphology to a spraying jet. Figure \ref{fig:5} presents the formation of the spraying jet with $\gamma=0.76$ and $\theta=1.23$. In contrast to the conical and umbrella-shaped jets, the lower surface of bubble 2 extends into the interior of bubble 1 during its expansion phase (frames 3-4). Subsequently, the upper surface of bubble 1 impacts the lower surface, initiating an annular collapse that emits a shock wave. This shock wave induces neck breakup at the elongated tip of bubble 2 (frames 5-7). The high-pressure stagnation point generated by the neck breakup drives the rapid upward acceleration of the liquid jet \citep{fan2024amplification}, culminating in the formation of a high-speed jet with spray at its tip. The spraying jet, being significantly faster, completes its trajectory from formation (initiated by neck breakup) to impact the upper surface of bubble 2 within the time interval of three frames. The relative measurement error is caused mainly by the limited temporal resolution, with the temporal uncertainty reaching up to one inter-frame interval and the resulting relative error in velocity as high as 50\% \citep{gonzalez2020jetting}.
The measured maximum velocity of this regime of jet can exceed 1000 m/s in the experiments. It is noteworthy that the tip of this jet is inherently unstable. Both experimental observations and numerical simulations reveal that the upper surface of bubble 2 is initially impacted by fine droplets, and then penetrated by the continuous jet. Here, we refer to the continuous jet as the uninterrupted liquid column behind the spraying droplets, characterized by a slower velocity compared to the unstable jet tip. The velocity of the continuous jet is particularly significant for practical applications that rely on coherent jet integrity. The experimentally measured velocity reaches approximately 375 ± 5.8 m/s, while the numerical simulation predicts a velocity of 417 m/s. Due to its narrower and faster continuous segment compared to conical and umbrella-shaped jets, this spraying jet is capable of achieving a propagation distance of up to 6.5 times the maximum bubble radius in this case. The mechanism of jet acceleration induced by neck breakup in two laser-induced bubbles has been demonstrated in \citet{fan2024amplification}. In this study, we similarly reproduced supersonic jets using centimeter-scale cavitation bubbles, but we observed a more detailed evolution process of the supersonic jet within bubble 2. We found that the jet is inherently unstable, characterized by a fragmented tip. The fragmented droplets sometimes follow an inclined trajectory, ultimately striking near the center of the upper surface of bubble 2, as shown in later discussions. Interestingly, although the trajectory of the spraying jet tip deviates from the symmetry axis of the tandem bubble system, the continuous jet quickly realigns itself, allowing it to continue penetrating and traveling along the axis.

\subsection {Penetration of piercing jets}
\begin{figure}
	\centerline{\includegraphics[width=1\textwidth]{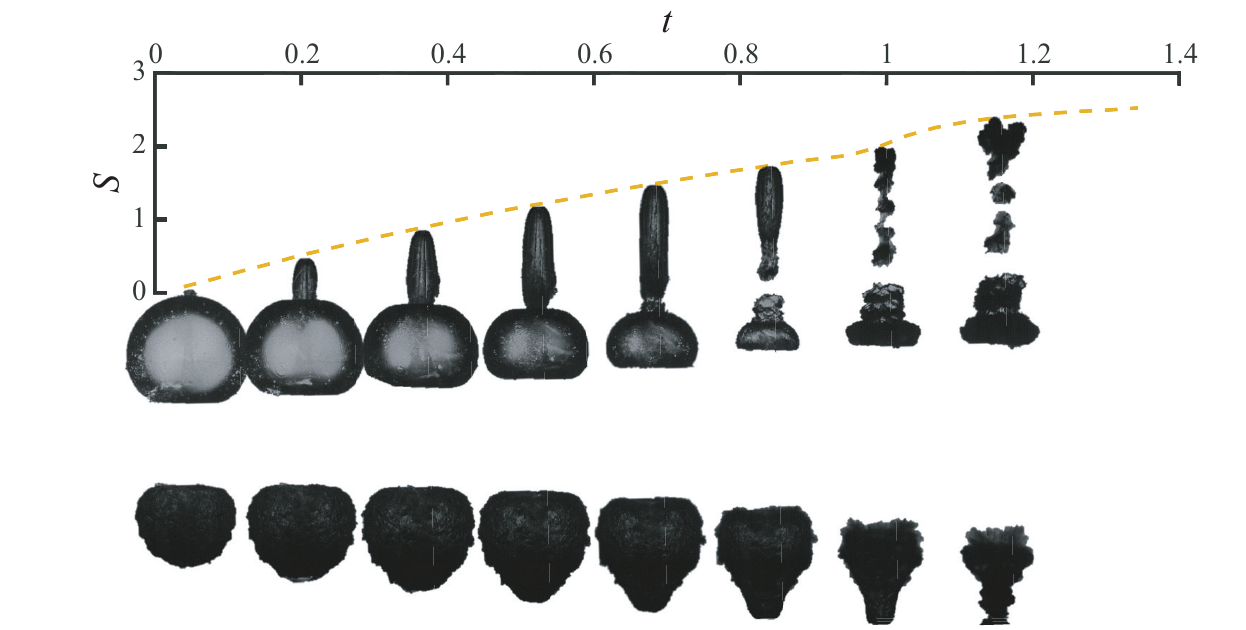}}
	\caption{The evolution of the gas cavity after the penetration of a conical jet. The curve depicts the temporal evolution of the distance $S$ penetrated by the jet tip in water. In this and subsequent figures, all variables without units are dimensionless. The time scale $ R_{\rm max} \sqrt{\rho / P_{\rm \infty}}$ is 2.25ms. The initial parameters are $\gamma=0.86$, $\theta=1.53$.}
	\label{fig:16}
\end{figure}

\begin{figure}
	\centerline{\includegraphics[width=1\textwidth]{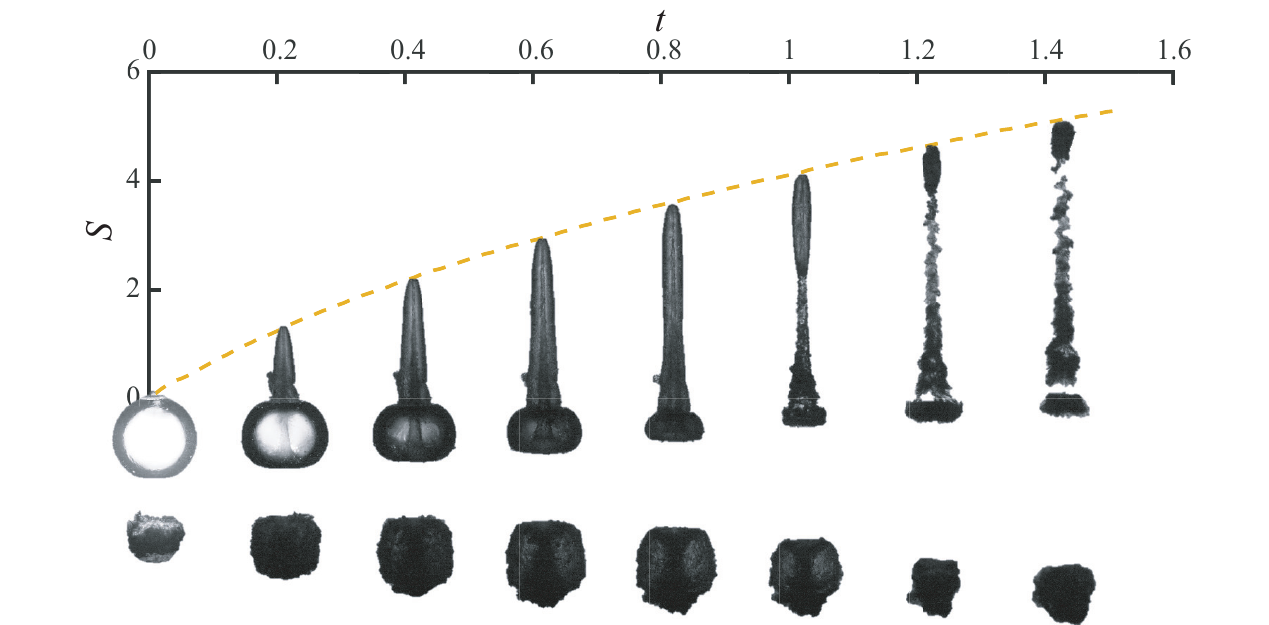}}
	\caption{The evolution of the gas cavity after the penetration of an umbrella-shaped jet. The curve depicts the temporal evolution of the distance $S$ penetrated by the jet tip in water. The time scale $ R_{\rm max} \sqrt{\rho / P_{\rm \infty}}$ is 2.25ms. The initial parameters are $\gamma=0.71$, $\theta=1.53$.}
	\label{fig:17}
\end{figure}

\begin{figure}\hspace{2mm}
	\centerline{\includegraphics[width=1.1\textwidth]{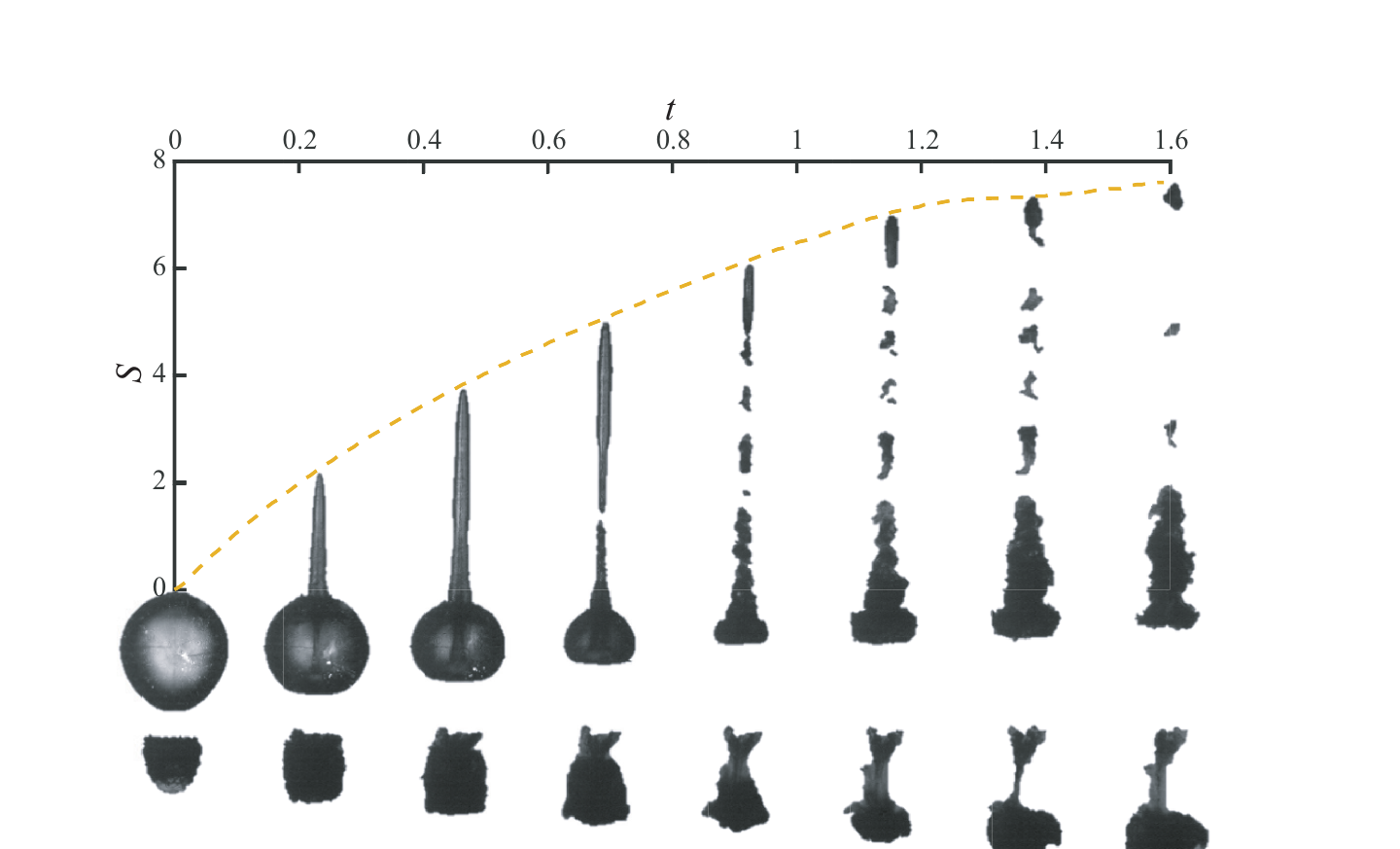}}
	\caption{The evolution of the gas cavity after the penetration of a spraying jet. The curve depicts the temporal evolution of the distance $S$ penetrated by the jet tip in water. The time scale $ R_{\rm max} \sqrt{\rho / P_{\rm \infty}}$ is 2.25ms. The initial parameters are $\gamma=0.85$, $\theta=0.95$.}
	\label{fig:18}
\end{figure}

After the piercing jets penetrate the opposite surface of bubble 2 and exit into the ambient liquid, we consistently observe that each jet is enveloped by a gas cavity during its propagation, a phenomenon similar to the gas cavity in water-entry problems, as shown in Figures \ref{fig:16}, \ref{fig:17}, and \ref{fig:18}. The time coordinate is normalized to zero at the critical moment of jet penetration, establishing a consistent reference for comparative analysis.
 
The evolution of the bubbles and the penetration process of a conical jet is illustrated in Figure \ref{fig:16}, where the curve depicts the temporal evolution of the distance penetrated by the jet tip. In this case, the conical jet retains a velocity of only 24 m/s after penetrating the bubble surface, with the jet tip advancing a distance of merely 2 times the maximum bubble radius before its collapse and fragmentation. Instants 1-2 reveal that the conical jet generates a continuous gas cavity trailing behind its tip upon penetrating the bubble surface. The cavity initially pinches off near the bubble (instants 3-5) and, after penetrating a stable distance (instants 5-6), disintegrates into bubble clusters (instant 7). As the velocity of the jet tip decreases, the cavity at the tip undergoes collapse followed by immediate expansion, resulting in a noticeable increase in the distance curve at instant 7. Owing to the absence of upstream jet replenishment, the cavity at the tip rapidly loses its forward momentum and gradually dissipates (instant 8). Hence, we only focus on the penetration distance at the instant of tip collapse, thereby excluding the subsequent cavity evolution from our study. 


Compared to the conical jet, the umbrella-shaped jet exhibits a higher velocity and retains a velocity of approximately 63 m/s after penetrating the bubble, which facilitates the formation of an extended cavity in the water (instants 2-4) and allows for a penetration distance exceeding 5 times the maximum bubble radius, as depicted in Figure \ref{fig:17}. Under this circumstance, bubble 2 undergoes significant elongation. The cavity adjacent to the toroidal bubble pinches off and collapses immediately after the toroidal collapse of bubble 2, followed by successive collapse propagating downstream along the cavity (instant 6). The cavity at the jet tip eventually collapses and its forward momentum gradually dissipates (instant 8).

The morphological evolution of the gas cavity after a spraying jet penetrates the bubble surface is shown in Figure \ref{fig:18}.  In comparison to conical and umbrella-shaped jets, the spraying jet demonstrates a significantly higher velocity after penetrating the bubble surface (108 m/s in the presented case), which results in the formation of a more elongated gas cavity, extending up to 6.5 times the maximum bubble radius. Bubble 2 undergoes significant elongation, resulting in markedly weaker secondary rebound compared to bubble 1 after collapse. A considerable amount of bubble 2's energy converts into jet kinetic energy, resulting in a weak collapse. Meanwhile, bubble 1's shockwave is shielded by bubble 2, thereby minimizing indiscriminate damage to the upper target injection site.

The cavity evolution after jet penetration resembles that produced by the water entry of a rigid body \citep{speirs2018water}. The subsequent breakup of these cavities is commonly characterized using Bond ($Bo = {\rho g D^2}/{\sigma}$) and Weber ($We = {\rho U^2 D}/{\sigma}$) numbers \citep{aristoff2009water,truscott2014water,speirs2018water}. Given that the tip radii span overlapping ranges across the three regimes (0.75-1.50 mm for the conical jet, 0.50-1.25 mm for the umbrella-shaped jet, and 0.25-0.75 mm for the high speed spraying jet), for these low–Bond number ($Bo \sim O(10^{-1}$-$10)$) cases, the cavity shape is dependent on the Weber number \citep{aristoff2009water,truscott2014water,speirs2018water}. At low Weber numbers, both conical jet and umbrella-shaped jet penetrate the bubble surface and undergo the shallow seal. In contrast, spraying jets, associated with higher $We$, form cavities where inertia force prevails, resulting in a deep seal under hydrostatic pressure influenced by bubble 2. It should be noted that a splash occurs after spraying jets penetrate the bubble surface and enter the water, as shown in frame 10 of Figure \ref{fig:5}. However, since these jets typically form when the bubble is at its maximum radius, a stage characterized by relatively low internal gas density, the associated pressure drop at the splash interface remains inadequate to drive closure. As expected, a deep seal rather than a surface seal is observed. The pinch-off types of cavity vary significantly across jet types and are accompanied by distinct flow structures corresponding to different jet tip morphologies, but they negligibly influence the maximum penetration distance, which is governed primarily by jet velocity. This study will focus on the  penetration capability of the jets in water, as detailed in \S{6}.

\section {The formation mechanisms of conical and umbrella-shaped jet}\label{4}

As previously demonstrated, bubble 2 develops an elongated tip under the influence of bubble 1. The high curvature at the elongated tip, coupled with the collapse of bubble 1, collectively govern the jet formation dynamics. This section employs BIM to temporally decouple the synergistic effects and systematically compare their influence on jet formation in bubble 2.

\subsection {Conical jet formation: the role of the pressure wave from bubble 1}\label{4.1}

As illustrated in \S{3.1}, the formation of the conical jet is initiated by the transient high pressure wave generated by the collapse of bubble 1, which accelerates the elongated tip of bubble 2 and drives its contraction into the bubble interior to form the liquid jet.  
One may argue that the high curvature at the tip of bubble 2 also contributes to the formation of a liquid jet \citep{lauterborn1982cavitation,tomita2002growth,sieber2022dynamics,zeng2024jetting}. However, whether the formation of the conical jet is driven by the high curvature or pressure wave remains to be elucidated.
\begin{figure}
	\centerline{\includegraphics[width=1\textwidth]{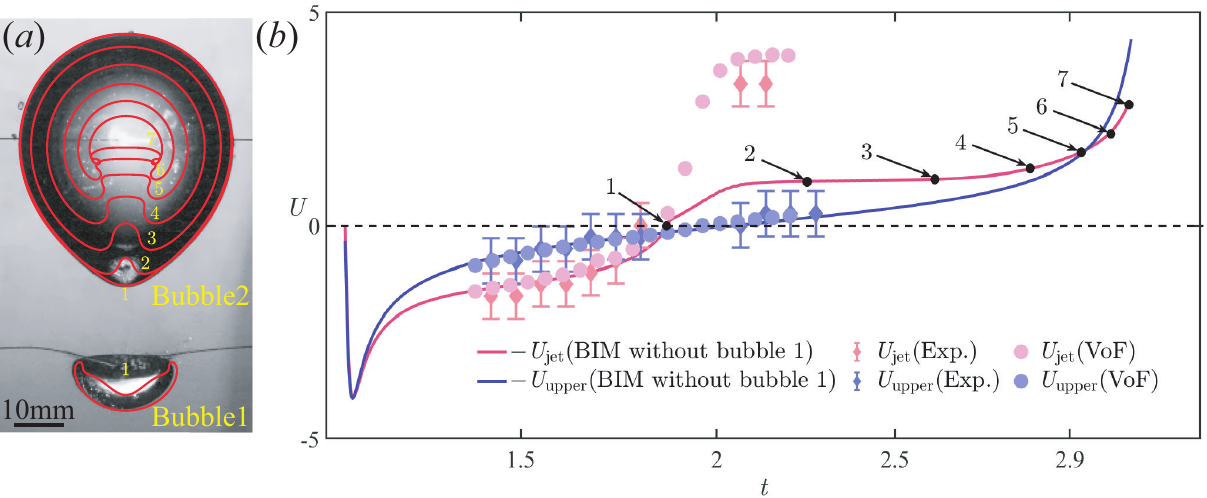}}
	\caption{
    Boundary integral simulation of the jet evolution of bubble 2. Bubble 1 is removed at the point when the downward velocity of bubble 2's lower tip reaches zero. ($a$) Morphological evolution of bubble 2. Instants 1-7 depict different snapshots during bubble evolution from BI simulation. ($b$) Time evolution of the velocity along the axis of symmetry of bubble 2. The red and blue solid lines represent the velocity of the jet and north pole of bubble 2 in the BI simulation, respectively. The solid diamonds with error bars represent the corresponding experimental data. The velocity is measured using three consecutive frames from high-speed imaging. The error is primarily attributed to spatial resolution, where the positional error corresponds to the length of one pixel. Temporal error arising from camera jitter occurs on the nanosecond scale and is therefore negligible. The solid circles represent the results obtained from the VoF simulation. The time scale $ R_{\rm max} \sqrt{\rho / P_{\rm \infty}}$ is 2.25ms. The initial parameters are $\gamma=0.98$, $\theta=1.10$.
}
	\label{fig:7}
\end{figure}

We first examine the role of the high curvature at the elongated tip of bubble 2 in driving the formation of the conical jet through BIM simulation. To eliminate the effect of the collapse of bubble 1 on bubble 2, bubble 1 is removed from the simulation when the downward velocity of bubble 2's lower tip reaches zero \citep{fong2009interactions,dadvand2012boundary,Peters2013,han2022interaction}. The subsequent morphological evolution and the time evolution of the velocity along the axis of symmetry of bubble 2 are presented in Figure \ref{fig:7}. Here, the red line represents the jet velocity, and the blue line represents the velocity at its north pole in Figure \ref{fig:7}($b$). Both velocities are defined as negative when the bubble surface expands outward and positive when it contracts inward to facilitate comparison. The experimental and VoF data, represented by solid diamond and square symbols, are also included. Instant 1 marks the moment when the expansion velocity at the lower tip of bubble 2 decreases to zero. At this point, bubble 2 expands into an egg-like shape under the influence of bubble 1, as shown in Figure \ref{fig:7}(\textit{a}). Bubble 1 is then removed from the simulation to eliminate the influence of its pressure wave. As expected, the region with a higher curvature on the bubble surface undergoes rapid contraction \citep{lauterborn1982cavitation,tomita2002growth}, contributing to jet formation at the lower tip of bubble 2. However, the simulated velocity and acceleration of the jet are significantly lower than experimental observations and VoF results in Figure \ref{fig:7}(\textit{b}). In BIM simulation, the upper surface of the bubble and the jet tip contract at similar velocities (instants 5-7). The continuous widening of the tip transforms the conical jet into a cylindrical shape, which differs from our experimental observations. We can conclude that while the high curvature of the bubble surface may contribute to the initial jet formation stage, it does not primarily govern the subsequent evolution of the conical jet.

\begin{figure}
	\centerline{\includegraphics[width=1\textwidth]{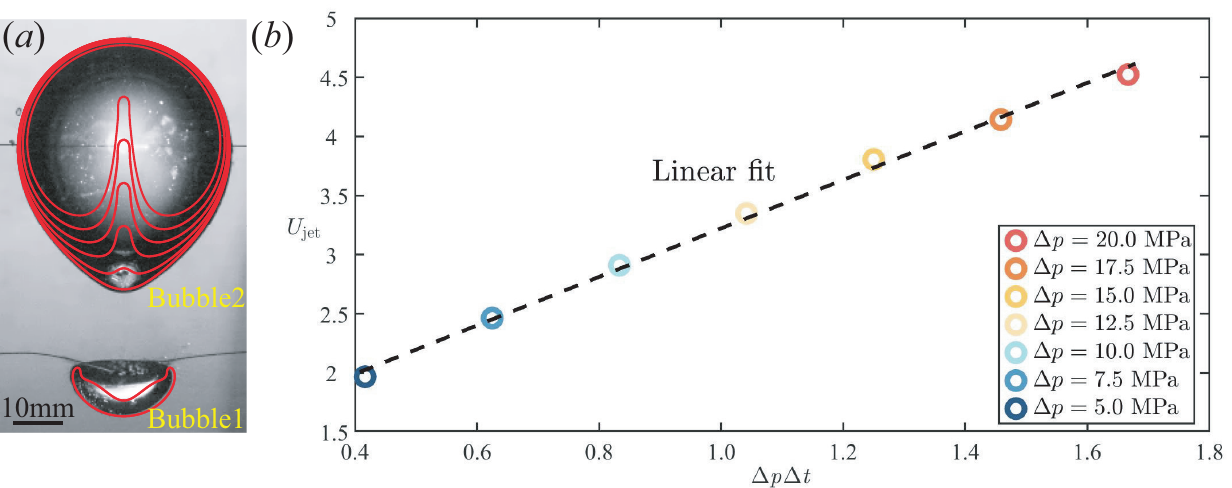}}
	\caption{Boundary integral simulation of the jet evolution of bubble 2. A pressure pulse with a duration of 20 \textmu s is applied within bubble 1. ($a$) Morphological evolution of bubble 2. The dimensionless time for the morphology of bubble 1 is 1.58, while the morphologies of bubble 2 from the outermost to the innermost correspond to dimensionless times of 1.58, 1.64, 1.73, 1.82, 1.93, and 2.04, respectively. ($b$) The maximum velocities of the conical jet of bubble 2 corresponding to different pressure pulses applied to bubble 1. The time scale $ R_{\rm max} \sqrt{\rho / P_{\rm \infty}}$ is 2.25ms. The initial parameters are $\gamma=0.98$, $\theta=1.10$.}
	\label{fig:8}
\end{figure}

Following \citet{ory2000growth} and \citet{Peters2013}, we adjust the pressure pulse from bubble 1 to investigate the role of the pressure wave in the formation of the conical jet. When the tip velocity of bubble 2 decreases to zero during its expansion phase, we hold the geometric profile of bubble 1 constant and apply a pressure pulse with a duration of 20 \textmu s while varying its amplitude. The duration of the pressure impulse generated by the collapse of bubble 1 is experimentally determined to be approximately 20 \textmu s through hydrophone measurements \citep{Cui2020}. The pressure amplitude is constrained within the range of 5–20 MPa \citep{Cui2020} to ensure the accelerated jet velocity remains within physically realistic limits. After the pressure pulse, the internal pressure of bubble 1 is set to hydrostatic pressure to eliminate any subsequent influence.

Figure \ref{fig:8} illustrates the tip evolution of bubble 2 under the impact of the pressure pulse from bubble 1. The temporal sequence of bubble 2’s morphology in Figure \ref{fig:8}(\textit{a}) shows a significant change compared to the simulation in Figure \ref{fig:7}(\textit{a}). The conical jet forms under the combined effects of tip curvature and the pressure wave from bubble 1, consistent with our experimental observations. As suggested by \citet{ory2000growth} and \citet{Peters2013}, when the interface is predominantly governed by the pressure pulse, the jet velocity scales with the product of pressure pulse amplitude and duration, expressed as $U_{\rm jet} \propto \Delta p \Delta t$. This relationship is well verified by our numerical simulation, as shown in Figure \ref{fig:8}(\textit{b}). In summary, we conclude that the relative contributions of surface curvature and pressure wave vary across different stages of jet development. A high-curvature surface is crucial for jet initiation, as regions with greater curvature undergo retraction first \citep{lauterborn1982cavitation,tomita2002growth}. However, the subsequent jet evolution is more governed by the pressure wave, which is corroborated by our finding that the velocity of the conical jet exhibits a positive correlation with the product of the pressure pulse amplitude and its duration. Therefore, the interplay of curvature and pressure wave is key to the formation of conical jets.

\subsection {Umbrella-shaped jet formation: bubble 1 collapses after jet formation}\label{4.2}

\begin{figure} \hspace{20mm}
	\centerline{\includegraphics[width=1\textwidth]{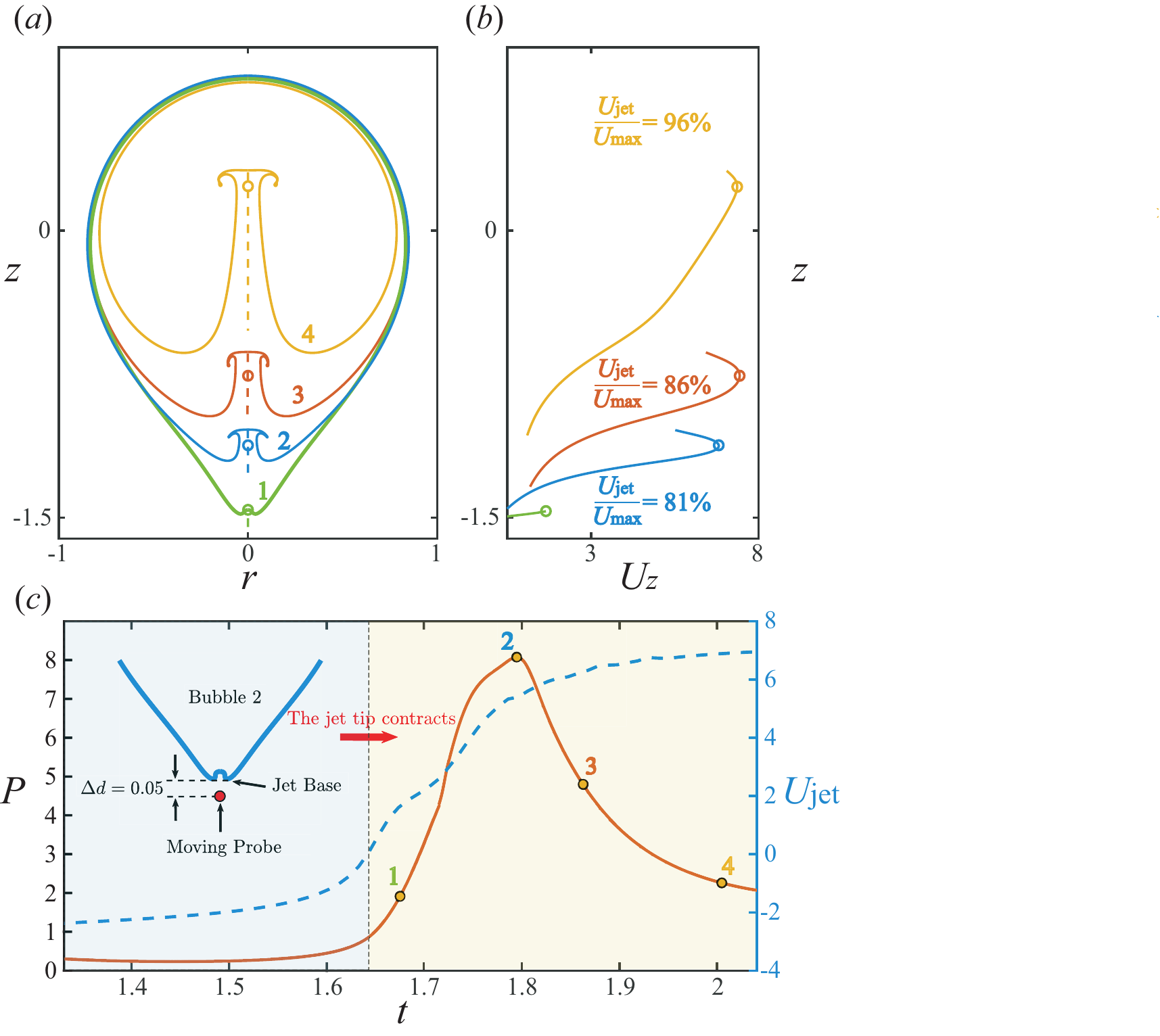}}
	\caption{Boundary integral simulation of the umbrella-shaped jet evolution of bubble 2. 
		($a$) Morphological evolution of bubble 2. The dashed lines indicate the position for velocity extraction.
		($b$) The evolution of axial velocity ($U_{\rm z}$) along the axis of symmetry
		in jet column of bubble 2. $U_{\rm jet}$ and $U_{\rm max}$ represent the velocity at the jet tip and the maximum velocity along the axis of symmetry in jet column, respectively.  
		The same color of solid lines and labels in ($a$) and ($b$) represents the same instants (1.68, 1.80, 1.86, 2.01). 
		The circles mark the location of the maximum velocity along the jet column. 
		($c$) Time evolution of pressure at a moving probe (0.05 dimensionless units below the jet base, solid line) and the time evolution of velocity at the jet tip (dashed line). The jet tip contracts at  $\textit{t}\approx 1.64$. The bubble profiles at times corresponding to labels \textrm {1}, \textrm {2}, \textrm {3}, and \textrm {4} are shown in panel (\textit {a}).
		The time scale $ R_{\rm max} \sqrt{\rho / P_{\rm \infty}}$ is 2.25ms. The initial parameters are $\gamma=0.83$, $\theta=1.10$.}
	\label{fig:9}
\end{figure}

We have provided preliminary insights into the formation of the umbrella-shaped jet in Figure \ref{fig:4}, characterized by a flattened leading edge and a fragmented liquid film around the jet tip. The umbrella-shaped jet resembles the mushroom-shaped structure observed when a liquid jet impinges against stagnant gas \citep{shinjo2010simulation}. Air resistance is identified as the primary driver of mushroom-shaped structure formation. Interestingly, our simulations using the BIM, based on potential flow theory, successfully reproduced a similar umbrella-shaped tip, as depicted in Figure \ref{fig:9}(\textit{a}). Given that the BIM does not account for air resistance and assumes uniform pressure within the bubble, this suggests that the air flow is not the primary cause of the umbrella-shaped structure formation. Through experimental observations, we find that the umbrella-shaped jet typically forms when the collapse of bubble 1 occurs after the initial jet formation stage of bubble 2. This finding demonstrates that the umbrella-shaped jet in the tandem bubbles system correlates with the flow structure within the water jet, which is governed by bubble-bubble coupling effects. This section will offer a more detailed analysis of the formation of the umbrella-shaped jet.

We present the evolution of axial velocity along the jet column from the BIM simulation in Figure \ref{fig:9}(\textit{b}). Meanwhile, the evolution of pressure below the jet base and the velocity of the jet tip are illustrated in Figure \ref{fig:9}(\textit{c}). The tip of the lower surface of bubble 2 contracts at $\textit{t}\approx 1.64$ (the black dashed line in Figure \ref{fig:9}(\textit{c})), accompanied by a short increase in velocity. During the initial formation of the jet, the peak velocity of the jet occurs at its tip driven by the local high curvature. However, due to the continuous increase in pressure at the jet base, which is induced by the focusing flow below bubble 2, the velocity of the upstream fluid (the fluid close to the jet base) gradually exceeds the tip velocity of the jet during its subsequent evolution. 
As the upstream fluid overtakes the jet tip, the liquid at the tip is displaced and spread outward, generating an umbrella-shaped structure, as shown by instant 2 in Figure \ref{fig:9}(\textit{a}). The jet tip experiences a noticeable reduction in acceleration after the collapse of bubble 1. This reduction results from the collapse-induced pressure accelerating the fluid near the jet base and widening the liquid jet column, thereby increasing its fluid inertia and consequently enhancing its resistance to acceleration. With the rebound of bubble 1, the pressure at the jet decreases gradually. The location of the maximum velocity of the jet gradually approaches the jet tip, resulting in a decrease in the velocity difference between the upstream fluid and the tip, as shown between instants 3 and 4 in Figure \ref{fig:9}(\textit{b}). A similar umbrella-shaped structure can also be observed in the downward Bjerknes bubble jet when the bubble is generated near a free surface under some circumstances \citep{supponen2015inner,koukouvinis2016simulation}. The free surface functionally emulates bubble 1, thereby inducing the characteristic egg-like deformation of the beneath bubble. As the high-curvature tip of bubble 2 contracts, a momentum-focusing effect occurs \citep{lauterborn1982cavitation,philipp1998cavitation, koukouvinis2016numerical,koukouvinis2016simulation}, causing the velocity of the trailing jet to exceed that of the jet tip, thereby generating a similar umbrella structure. These results demonstrate that umbrella-shaped jet formation can be initiated solely by pressure increase at the jet base during its evolution, with pressure waves being non-essential.

\begin{figure}
	\centerline{\includegraphics[width=1.2\textwidth]{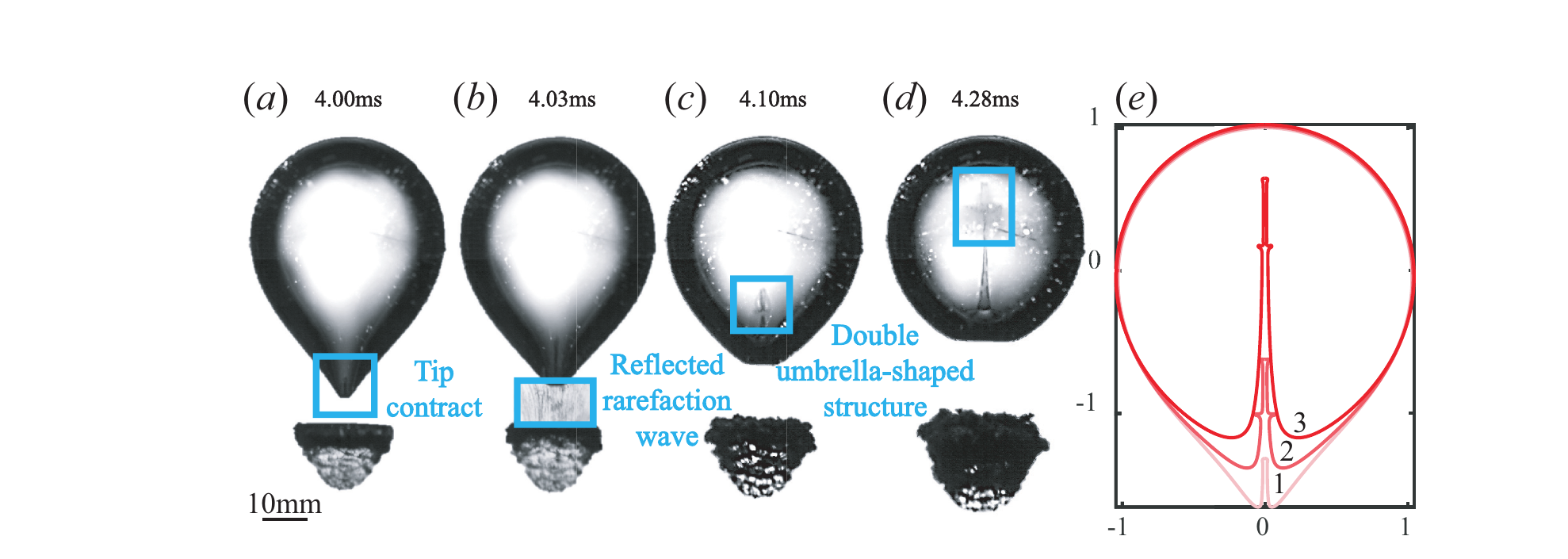}}
	\vspace{-10pt}
	\caption{Evolution of the double umbrella-shaped jet. In panels (\textit{a}) to (\textit{d}), the lower bubble is bubble 1 and another is bubble 2. The blue solid line boxes indicate tip contract, reflected rarefaction wave and double umbrella-shaped structure. Panel (\textit{e}) presents the boundary integral simulation of the double umbrella-shaped jet evolution of bubble 2. A 20 \textmu s pressure pulse of 60 MPa is applied within bubble 1 at instant 1. The dimensionless time corresponding to Instants 1, 2, and 3 are 1.76, 1.80, and 1.88 respectively. The time scale $ R_{\rm max} \sqrt{\rho / P_{\rm \infty}}$ is 2.25ms. The initial parameters are $\gamma=0.83$, $\theta=1.06$.}
	\label{fig:10}
\end{figure}

Interestingly, we observed the formation of a double-umbrella shaped jet under specific conditions in our system, as illustrated in Figure \ref{fig:10}. In panel (\textit{a}), the tip of bubble 2 begins to contract inward. A cluster of cavitation bubbles in panel (\textit{b}) marks the complete collapse of bubble 1, which triggers the emission of the pressure waves. Following this shock, two umbrella-shaped structures can be observed simultaneously, one at the tip and another in the middle of the jet, as illustrated in panels (\textit{c}) and (\textit{d}). The first umbrella-shaped structure originates from flow focusing, demonstrating that pressure waves are not essential for its formation. Conversely, the second structure develops from the liquid column that was widened by pressure waves emitted during bubble 1's collapse. This double umbrella-shaped jet aligns with findings by \citet{geschner2001investigation} and \citet{srinivasan2011modeling}, who demonstrated that imposing a pressure excitation after one umbrella structure on a liquid jet led to the formation of the second umbrella-shaped structure (see Figure 3 in \citet*{geschner2001investigation} ).

Through the BI simulation, we reproduced the evolution of this distinct jet, as shown in panel (\textit{e}). In this simulation, the geometric profile of bubble 1 is held constant when bubble 1 collapses to its minimum volume. At this moment (instant 1), the internal pressure of bubble 1 is modulated into a transient pressure pulse (duration of 20 µs, amplitude of 60 MPa), inducing a transient shock at the jet base of bubble 2. After the pressure pulse, the internal pressure of bubble 1 is set to hydrostatic pressure to eliminate any subsequent influence. Following the pressure pulse, an axial protrusion emerges at the jet midsection (instant 2), ultimately developing into a secondary umbrella-shaped structure (instant 3) that matches experimental observations. This further clarifies two critical aspects of the umbrella-shaped jet formation. First, the bubble has to develop a high-curvature tip to initiate the tip contraction. Second, it is the pressure increase at the jet base of bubble 2 (resulting from the focusing flow or pressure wave) that drives the formation of the umbrella-shaped jet. Meanwhile, we systematically vary the initial gas density parameters in the equation of state ($\rho_{\rm g0}$ = 0.26, 1.29, and 6.45 kg/m³) to represent different levels of gas content inside the bubble (the corresponding numerical results are not included here). The results indicate that gas density has only a minor influence on jet velocity (within 0.2\%), and the tip morphology of the umbrella-shaped jet remains generally consistent. These simulations demonstrate that the jet structure originates primarily from the liquid flow and is largely independent of the gas flow inside the bubble. This finding advances our fundamental understanding of how flow structure governs the tip morphology of jets.

\begin{figure}
	\centerline{\includegraphics[width=1\textwidth]{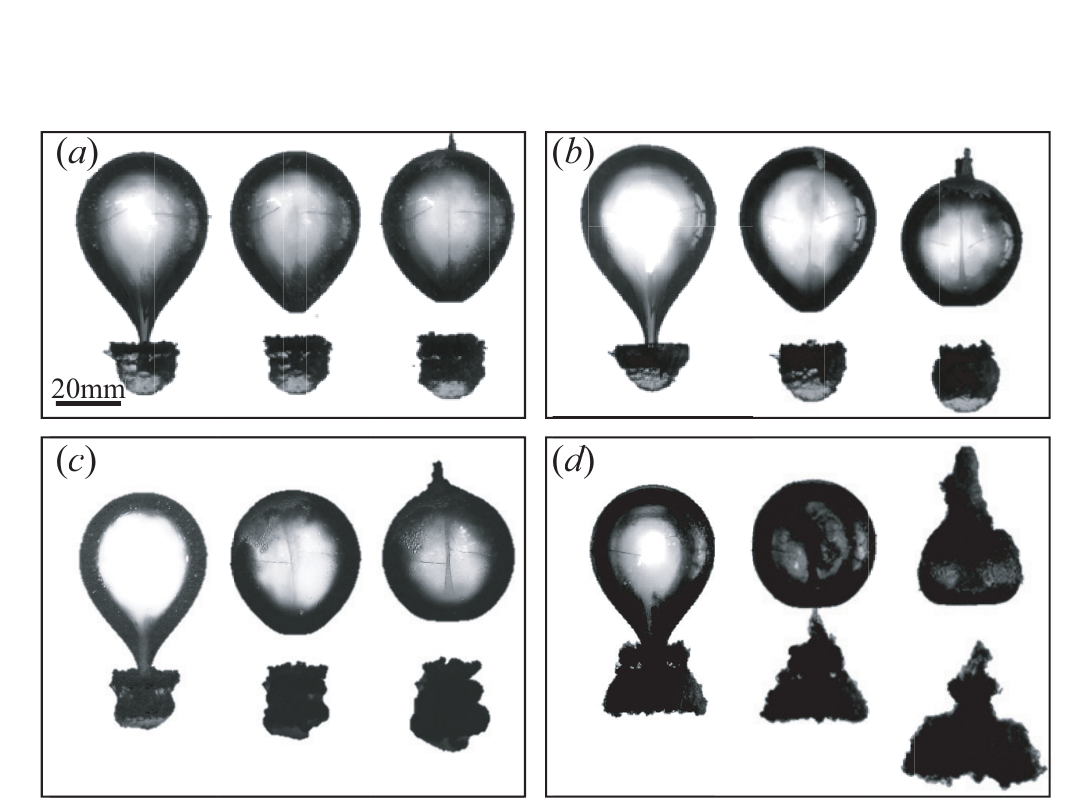}}
	\caption{Four types of spraying jets: (\textit{a}) Needle-like spraying jet. The initial parameters are $\gamma=0.76$, $\theta=1.25$, and the corresponding times for each frame are 4.14 ms, 4.19 ms, and 4.25 ms. (\textit{b}) Inclined spraying jet. The initial parameters are $\gamma=0.77$, $\theta=1.33$, and the corresponding times for each frame are 4.25 ms, 4.36 ms, and 4.53 ms. (\textit{c}) Inclined sheet-like spraying jet. The initial parameters are $\gamma=0.72$, $\theta=1.35$, and the corresponding times for each frame are 4.19 ms, 4.32 ms, and 4.49 ms. (\textit{d}) Mist-like spraying jet. The initial parameters are $\gamma=0.7$, $\theta=1.21$, and the corresponding times for each frame are 4.12 ms, 4.54 ms, and 5.58 ms.}
	\label{fig:11}
\end{figure}

\section{Dynamics and transition of spraying jets: from needle-like to mist-like structures}\label{5}
With a further decrease in $\gamma$, the location of pressure wave impingement shifts from the bottom point to the neck region of bubble 2's elongated tip, inducing the neck breakup that generates spraying jets. This process accelerates the jet velocity to exceed the sound speed of air. Four types of spraying jets are observed in the strong interaction between two tandem bubbles, as given in Figure \ref{fig:11}. The tip of the jet may fragment into small liquid droplets or form inclined/film-like sprays (Figure \ref{fig:11}(\textit{a})-(\textit{c})), which is related to the neck breakup of the elongated bubble 2. A further decrease in $\gamma$ induces bubble coalescence, which attenuates the pressure wave intensity from bubble 1. The bubble coalescence induces another spraying mode, namely mist-like spraying jet, which features a broader liquid jet with a splashing tip as shown in Figure \ref{fig:11}(\textit{d}).

 \subsection {Spraying jet caused by neck breakup}
\begin{figure} \hspace{9mm}
\centerline{\includegraphics[width=1\textwidth]{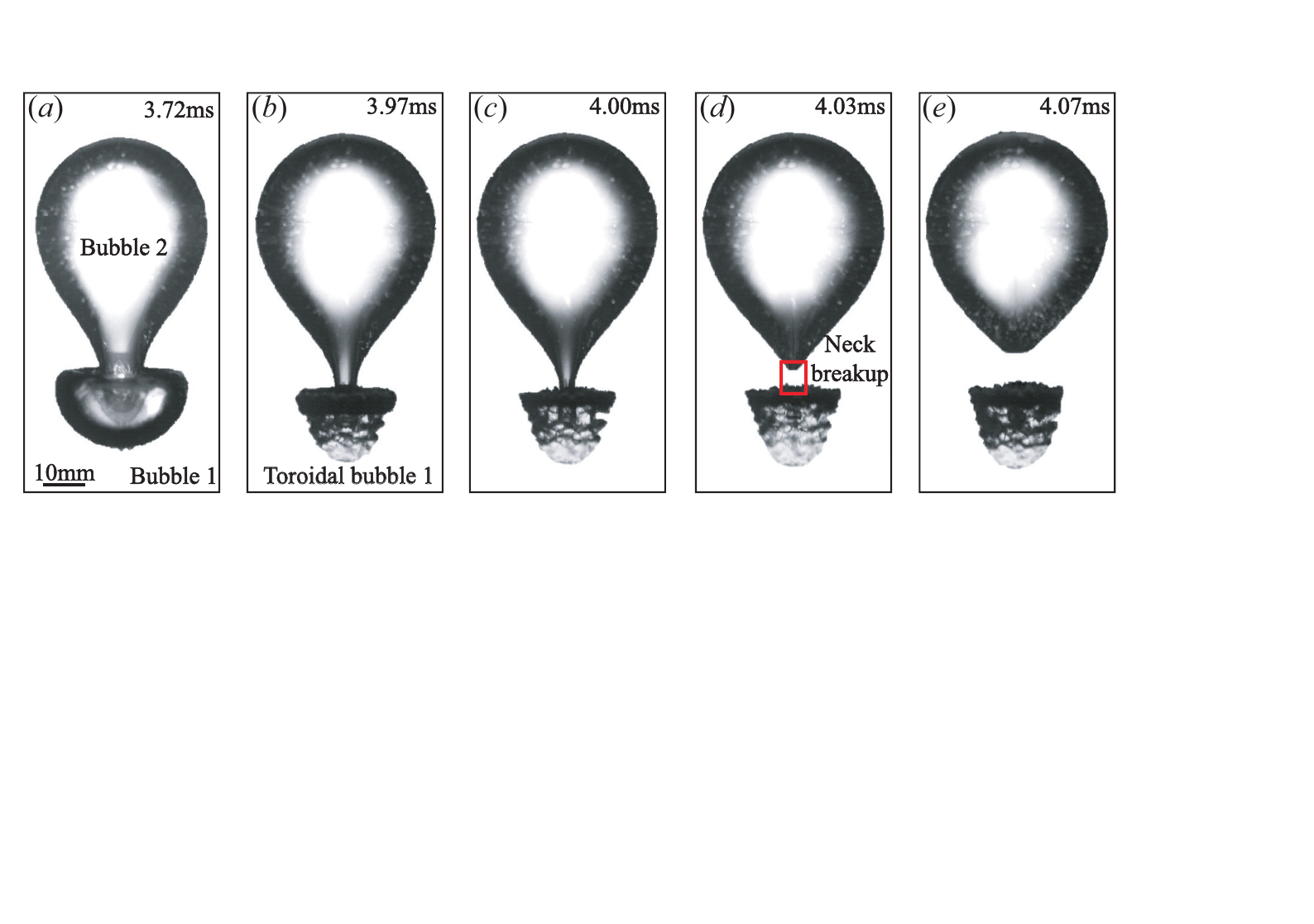}}
\vspace{-130pt}
  \caption{Evolution details of neck breakup process of bubble 2 following the impingement of bubble 1's jet on the lower surface. The initial parameters are $\gamma=0.79$, $\theta=1.11$. }
\label{fig:12}
\end{figure}

Figure \ref{fig:12} illustrates the detailed evolution of the neck breakup process of bubble 2. The tip of bubble 2 goes deeply into bubble 1 without coalescence during the contraction of bubble 1 (panel (\textit{a})). Consequently, the high pressure generated by the collapse of bubble 1 acts directly on the neck of the elongated tip of bubble 2 (panel (\textit{b})), initiating the neck breakup (panels (\textit{c})). Driven by the high pressure generated by the flow focusing during the neck breakup, the supersonic jet in centimeter-scale tandem bubbles can achieve velocities exceeding 1000 m/s (panel (\textit{d})), comparable to those in millimeter-scale experiments \citep{fan2024amplification}. Although the spraying jet accelerated by neck breakup is very fast, we observed that the tip of the spraying jet is always fragmented into mist droplets, as illustrated in Figure \ref{fig:11}(\textit{a})-(\textit{c}). In these experiments, the droplets first impact the bubble surface,  forming a cluster of tiny craters, while the continuous jet subsequently penetrates the upper surface. This behavior may be explained by the Worthington jet drop-off mechanism, as illustrated by \citet{gordillo2010generation}. Following the neck breakup, the jet tip attains significantly high kinetic energy and breaks up due to the capillary deceleration of the liquid. The detached droplets are then further fragmented into smaller droplets due to instability, a process known as secondary atomization \citep{guildenbecher2009secondary,choi2022analysis,huck2022spray}. This explains the fine droplet cloud observed at the tip of the spraying jets in the experiments.

Through extensive experiments, we have observed that supersonic jets with straight needle-like tips are relatively rare, while inclined jets (Figure \ref{fig:11}(\textit{b})) or inclined sheet-like jets (Figure \ref{fig:11}(\textit{c})) are significantly more prevalent. We attribute this prevalence to the inherent instability of the neck breakup process \citep{gekle2010generation}. Any initial perturbations during the breakup process can cause asymmetric impingement and non-singular collapse of the neck \citep{burton2005scaling, keim2006breakup, enriquez2012collapse}. 
Even minor initial perturbations persist throughout the neck breakup process due to memory effects \citep{keim2006breakup}. As a result, initial azimuthal asymmetries during neck breakup can easily affect the morphology and direction of the jet in the subsequent process, forming inclined spraying jets in Figure \ref{fig:11}(\textit{b}). \citet{schmidt2009memory} and \citet{enriquez2012collapse} revealed that when the initial elliptic cavity is nearly circular, the neck does not collapse into an infinitesimal singularity but instead impinges from two opposing sides, leading to the sheet-like jet, which closely resembles the jet observed in Figure \ref{fig:11}(\textit{c}).  

Although the tip of the spraying jet is fragmented into droplets due to instability, the subsequent continuous jet remains highly stable. The jet travels at a significantly high velocity with excellent directionality, well aligned with the axis of the two tandem bubbles. The good stability of jet direction is maintained due to the support by the jet base instead of the stagnation point of the neck breakup \citep{gekle2010generation}. Further details on the penetration capability of this jet in liquid will be provided in \S{6} and \S{7}.

\begin{figure}
	
\vspace*{0mm}	
	\centerline{\includegraphics[trim={0pt 0 0pt 180pt}, 
		clip, width=1\textwidth]{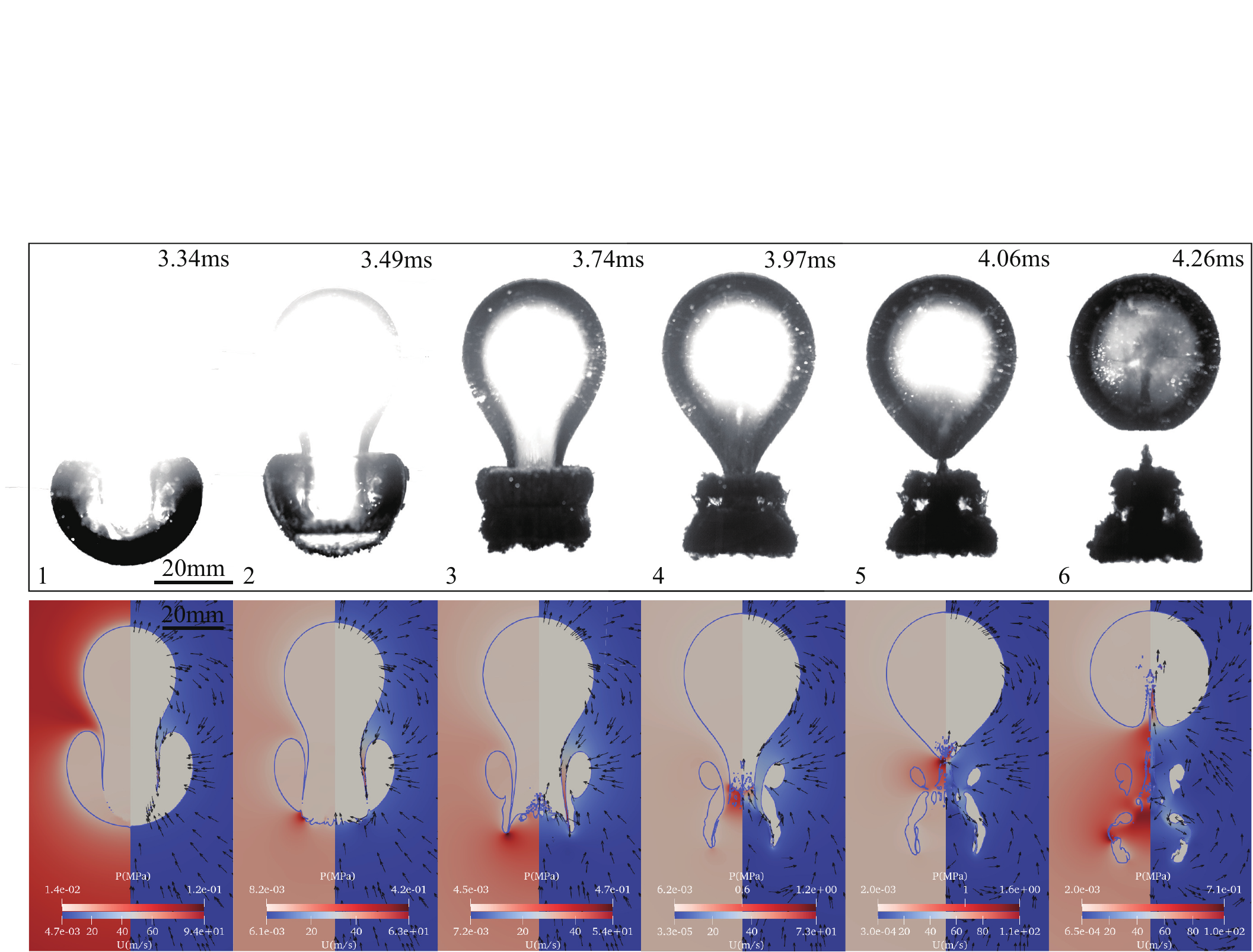}}
	\caption{Dynamics of mist-like spraying jet from anti-phase bubble pair: comparison between the experiment (odd rows) and numerical simulation (even rows). The dimensionless time corresponding to the numerical results are 1.38, 1.44, 1.56, 1.67, 1.74, and 1.87, respectively. The time scale $ R_{\rm max} \sqrt{\rho / P_{\rm \infty}}$ is 2.25ms. The initial parameters are $\gamma=0.65$, $\theta=1.19$.}
	\label{fig:15}
\end{figure}
\begin{figure}
	\centerline{\includegraphics[width=1\textwidth]{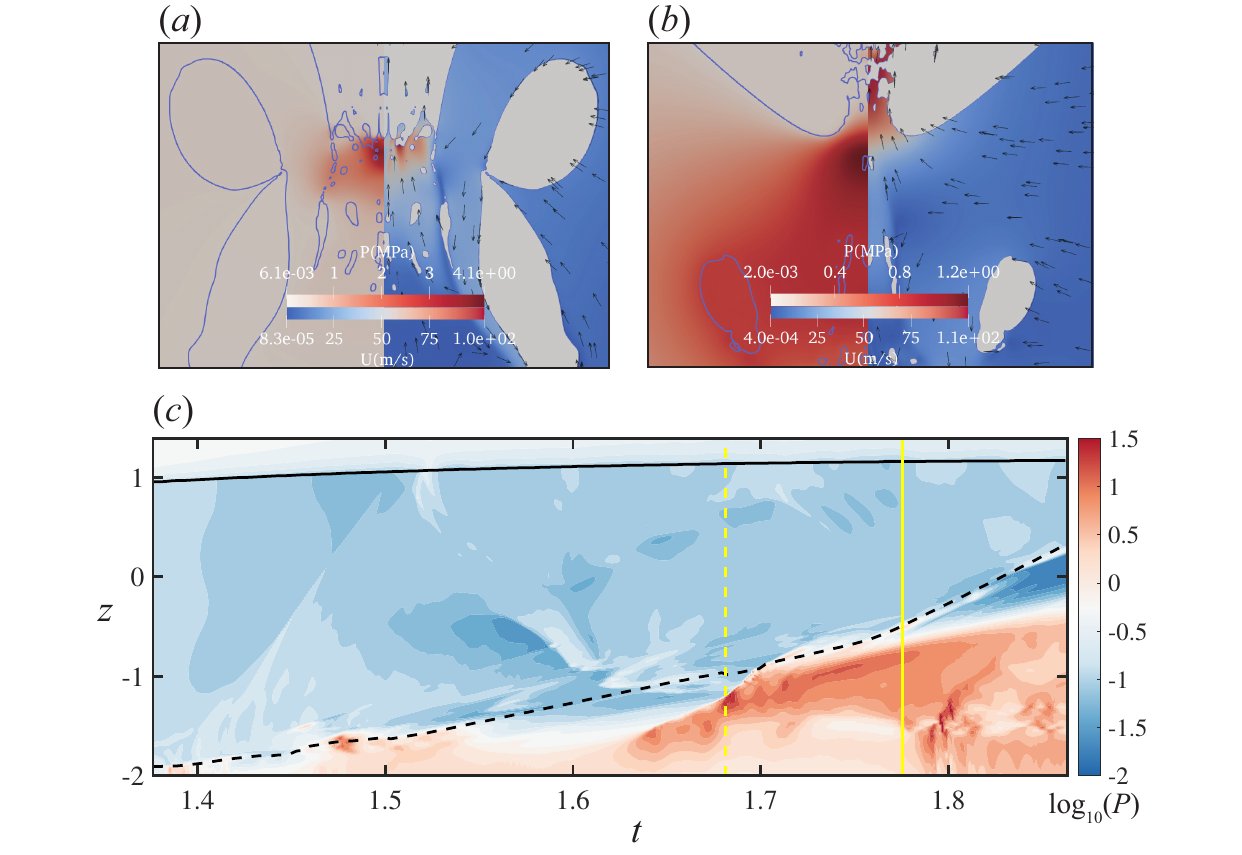}}
	\caption{
	The evolution of the mist-like spraying jet shown in Figure \ref{fig:15}. 
	($a$-$b$) Zoom-in details of the flow field at $t={\rm 1.68, 1.78}$. The small bubbles below the jet base undergo collapse and generate localized high-pressure regions.
    ($c$) Time-space map of pressure along the axis of symmetry of the two tandem bubbles. The black solid and dashed lines represent the axial position of the upper surface and jet tip of bubble 2, respectively. The yellow dashed and solid lines correspond to two typical moments in ($a$) and ($b$), respectively. The time scale $ R_{\rm max} \sqrt{\rho / P_{\rm \infty}}$ is 2.25ms.}
	\label{fig:14}
\end{figure}

 \subsection {Spraying jet caused by bubble coalescence}

The mist-like spraying jet formation involves deep penetration of bubble 2's elongated tip into bubble 1's interior, making experimental observation challenging. We therefore employ OpenFOAM simulations to complement the investigation of this jet evolution process. The overall evolution of the mist-like spraying jet is illustrated in Figure \ref{fig:15}, with velocity and pressure fields provided. The elongated tip of bubble 2 coalesces with the surface of bubble 1 due to the small $\gamma$ (instant 1). At instants 2-3, the tip of bubble 2 forms a fragmented jet during its contraction process, and the subsequent evolution of the fragmented jet leads to the formation of a cluster of dispersed tiny bubbles (instant 4). The dispersed tiny bubbles are drawn by bubble 2 due to its continuous contraction (instant 5). The fragmented liquid jet propagates further into the bubble interior, drawing a trailing gas tail at its base as visualized (instants 5-6). The jet exhibits a fragmented, radially splashing morphology (instant 6), which is quite different from the needle-like spraying jets. These jet evolution details as shown above prove experimentally inaccessible since bubble 2’s elongated tip remains either obscured by bubble 1 at instants 1-3 or manifests as a misty gas cluster at instants 4-5. 
Although we employ axisymmetric simulations rather than computationally intensive 3D modeling, the numerical results successfully capture the dominant physical processes in the experiment, thereby complementing the limitations of experimental observations and providing critical insights into the formation of mist-like spraying jets.

Enlarged views of the flow field surrounding the tiny bubbles at $t={\rm 1.68, 1.78}$ are presented in Figure \ref{fig:14}($a$)-($b$), respectively. These small bubbles undergo collapse and generate localized high-pressure waves to accelerate the fragmented jet. 
Additionally, the time-space map of the pressure along the symmetry axis during the formation and development of the jet is presented in Figure \ref{fig:14}(\textit{c}) to offer a more comprehensive perspective on the jet evolution. The black solid and dashed lines represent the axial position of the upper surface and jet tip of bubble 2, respectively. It can be seen that the upper surface of bubble 2 shows negligible changes during jet development. The yellow dashed and solid lines correspond to two typical moments, which are associated with ($a$) and ($b$), respectively. The intermittent collapse of these bubbles induces continuous pressure fluctuations at the jet base during the time interval $t=1.6$-$1.8$. Subject to these high-pressures, the fragmented jet continues to accelerate, and its instability is further amplified, leading to radial splashing (instants 5-6 in Figure \ref{fig:15}).

In contrast to the localized high-pressure stagnation point generated by neck breakup, the collapse of the dispersed tiny bubbles induces continuous high-pressure fluctuations that accelerate broader liquid jet formation. This results in a jet velocity substantially lower than that of needle-like spraying jets. From the perspective of needle-free injection applications, such mist-like jets, characterized by relatively low velocities and splashing tips, are undesirable and should be avoided.

\section {Jet penetration}\label{6}

\subsection {Preliminary remarks}

\begin{figure}
	\centerline{\includegraphics[width=1.1\textwidth]{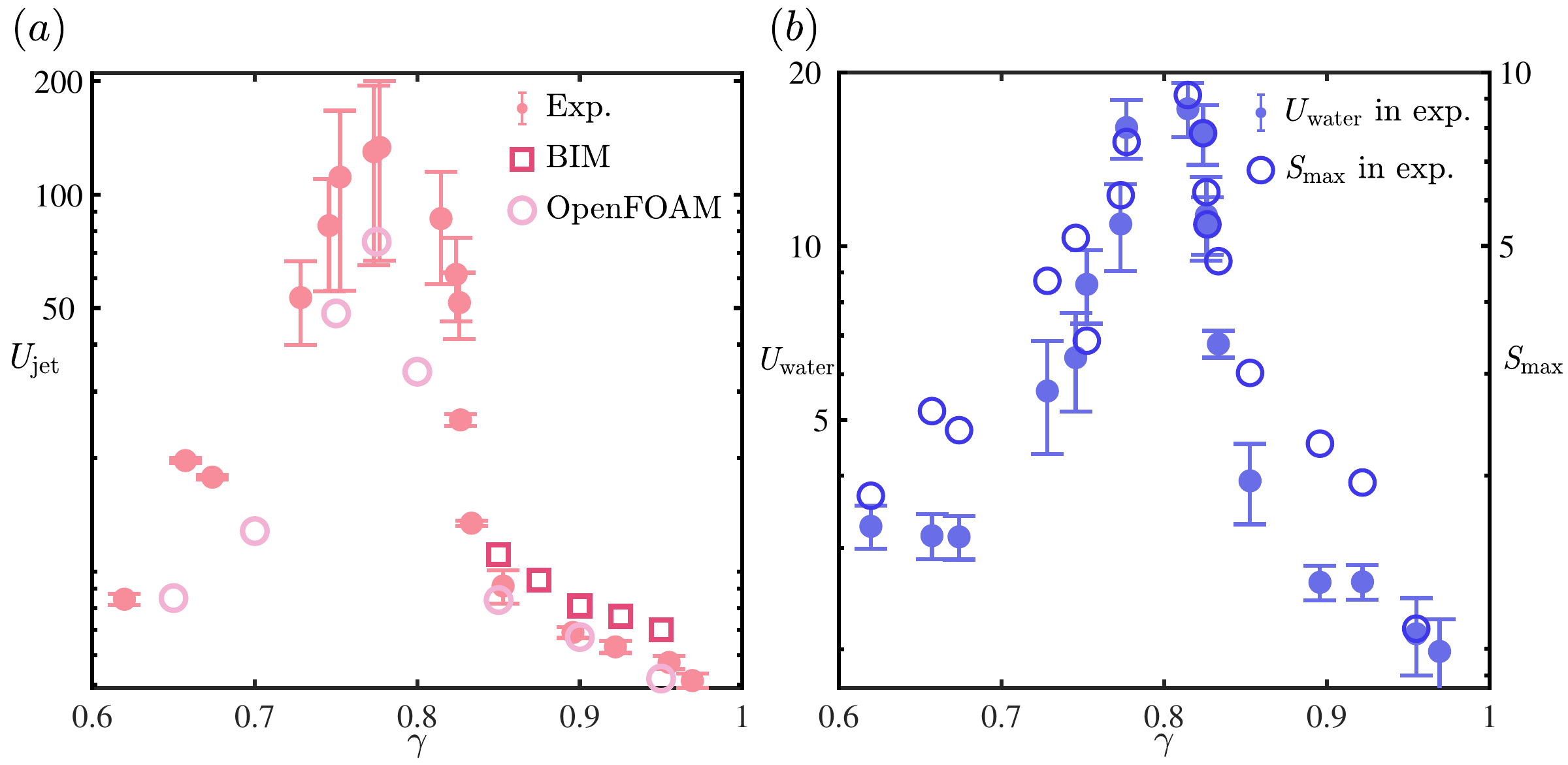}}
	\caption{Dependence of the velocity and maximum penetration distance of piercing jets on $\gamma$ at a fixed $\theta = 1$. The velocity is measured using several consecutive frames from high-speed imaging. The error is primarily attributed to spatial resolution, where the positional error corresponds to the length of one pixel. For the fastest jets (spraying jets), the experimentally measured velocity represents a lower bound of the real velocity due to the inherent temporal resolution limitations of high-speed imaging. The jet completes its trajectory from formation (initiated by neck breakup) to impact on the bubble surface within three frames, resulting in a maximum relative uncertainty as high as 50\% \citep{gonzalez2020jetting}. ($a$) Dependence of the internal velocity $U_{\rm jet}$ of piercing jets on $\gamma$. The solid symbols represent experimental data. The hollow symbols denote numerical results obtained from BIM and OpenFOAM. $U_{\rm jet}$ is the maximum velocity of the jet in bubble 2. ($b$) Dependence of the external velocity $U_{\rm water}$ and maximum penetration distance ${S_{\rm max}}$ of piercing jets on $\gamma$. $U_{\rm water}$ represents the velocity of the jet as it enters the liquid after penetration and ${S_{\rm max}}$ denotes the maximum penetration distance by the jet in the liquid at the moment of cavity protrusion collapse at its tip. All variables in the figure are dimensionless. The velocity scale $ \sqrt{P_{\rm \infty} / \rho}$ is 10 m/s.}
	\label{fig:6}
\end{figure}

The evolution of piercing jets within bubble 2 exhibits notable velocity transitions across three distinct regimes, with maximum velocities exceeding 1200 m/s in the spraying jet regime. 
However, the piercing jet experiences velocity attenuation after penetrating the bubble surface. Particularly for spraying jets, while the high-speed fragmented droplets at the jet tip can only disturb the bubble interface, the jets exhibit pronounced velocity differences between their internal (within bubble 2) and external (after penetrating bubble 2) velocities.
To further elucidate the penetration performance of these three jet regimes, we present the dependence of jet velocity and penetration distance on the initiation bubble distance $\gamma$ in Figure \ref{fig:6}. Here, the internal velocity $U_{\rm jet}$ is the maximum velocity of the jet in bubble 2, the external velocity $U_{\rm water}$ represents the velocity of the jet as it enters the liquid after penetration and ${S_{\rm max}}$ denotes the maximum penetration distance by the jet in the liquid at the moment of cavity protrusion collapse at its tip. The relative initiation bubble time difference, $\theta$, is held constant in both experiments and numerical simulations. The jet velocities obtained from the BIM and VoF show good agreement with experimental results in Figure \ref{fig:6}($\it a$). Due to its limitations in handling mesh topology on bubble surfaces, the BIM is used exclusively to simulate scenarios involving conical and umbrella-shaped jets.

As depicted in Figure \ref{fig:6}($\it a$), a distinct transition is observed with decreasing $\gamma$: the jet morphology evolves sequentially from a conical jet to an umbrella-shaped jet and ultimately to a spraying jet. Generally, the internal velocity $U_{\rm jet}$ exhibits a non-monotonic dependence on $\gamma$, first increasing then decreasing as $\gamma$ is reduced. When the value of $\gamma$ decreases to approximately 0.77, $U_{\rm jet}$ abruptly increases to exceed 120, corresponding to a dimensional value of 1200 m/s, as shown in Figure \ref{fig:appendix} provided in the Appendix. This increase is attributed to the extreme acceleration caused by the neck breakup at the elongated tip of bubble 2, which closely aligns with the velocity reported by \citet{fan2024amplification}. Further reducing the initial bubble-bubble distance leads to the coalescence between the two bubbles and contributes to the rapid decrease in $U_{\rm jet}$. The strong correlation between maximum penetration distance $S_{\rm max}$ and external jet velocity $U_{\rm water}$ in Figure \ref{fig:6}($\it b$) suggests that penetration performance is dominantly controlled by the external jet velocity $U_{\rm water}$ rather than the internal jet velocity $U_{\rm jet}$. Particularly for needle-free injection applications, the external jet penetration characteristics constitute the primary focus of this investigation.

\subsection {Penetration performance of piercing jets}

To describe the piercing jet dynamics after entering the liquid, we develop a simplified model by conceptualizing the jet as a liquid-bullet with initial velocity $U_{\rm water}$ and effective mass $m$. When such a liquid-bullet propagates through water, the entrained cavity envelops its main body, leaving only the tip in contact with the surrounding liquid. Consequently, the dominant drag force $F$ arises primarily from hydrodynamic pressure at the advancing tip, expressed as $F = -C_{\rm D} \rho A u^2$, where $\rho$ denotes the water density, $A$ denotes the jet's cross-sectional tip area (the cross-sectional area of the liquid phase at the cavity tip), and $u$ is the jet velocity. The drag coefficient $C_{\rm D}$ (primarily determined by the tip geometry) is taken as $C_{\rm D} = 0.5$ \citep{soh2005entrainment,mallick2014study,kuwata2021flow} throughout this study. Neglecting viscous and surface tension effects, the jet's momentum equation reduces to:
\begin{equation}
m \frac{\mathrm{d}u}{\mathrm{d}t} = -C_{\rm D} \rho A u^2,
\label{eq:momentum}
\end{equation}
where $m$ represents the effective jet mass. By separating variables and integrating equation (\ref{eq:momentum}), we obtain $u(t) = {U_{\rm water}}/{(1 + K U_{\rm water} t)}$, where $K = {C_{\rm D} \rho A}/{m}$. Thus, the penetration $S$ can be expressed 

\begin{equation}
S(t) = \int_0^t u(t) \, \mathrm{d}t = \frac{1}{K} \ln(1 + K U_{\rm water} t).
\label{eq:penetration_depth}
\end{equation}

\begin{figure}
	\centerline{\includegraphics[width=0.9\textwidth]{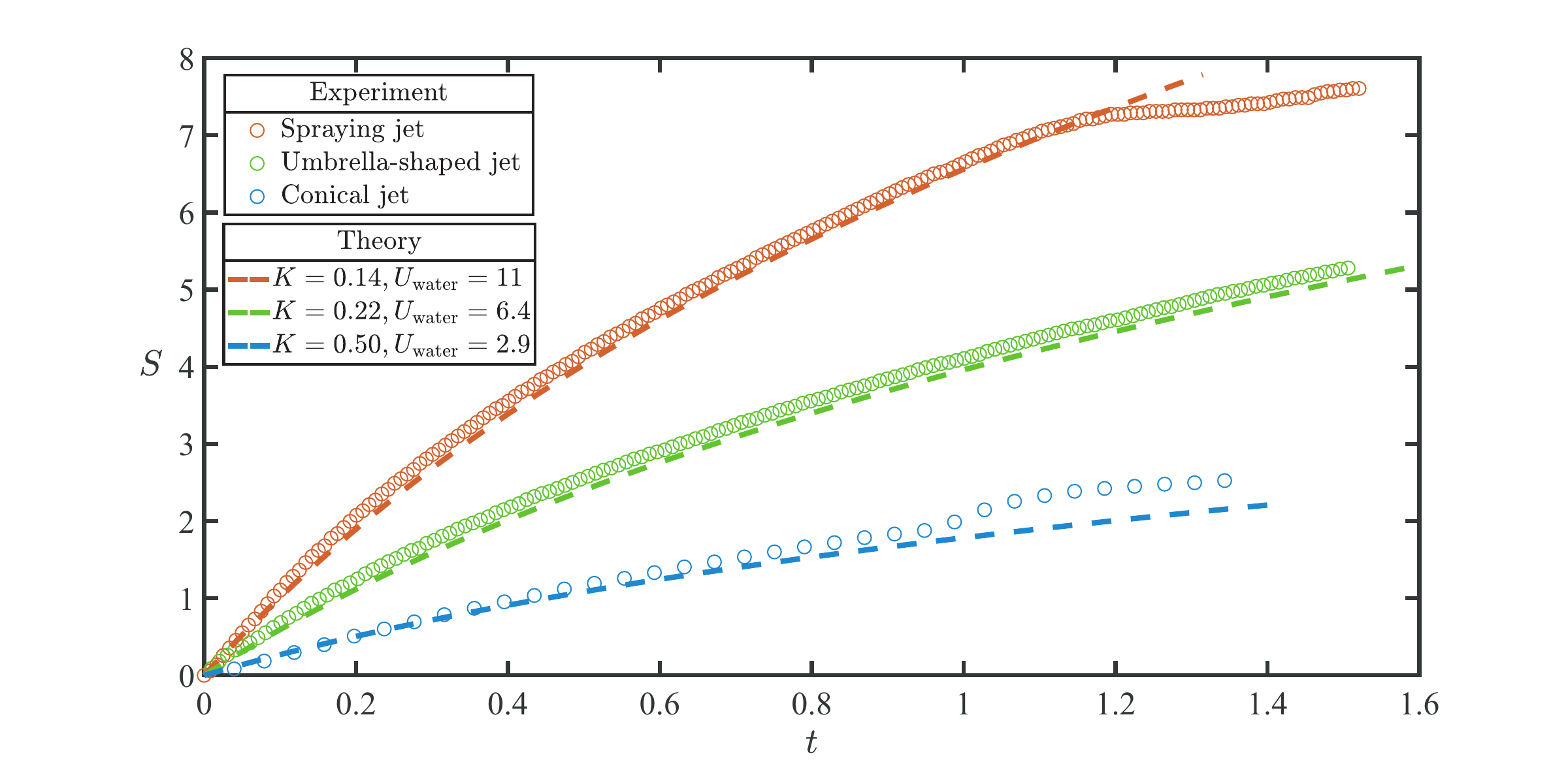}}
	\caption{Temporal curves of the penetration distance $S$ for three regimes of piercing jets. The hollow symbols denote experimental results and the dashed lines represent predictions of the liquid-bullet model given by equation (\ref{eq:penetration_depth}) for different values of $K$ and $U_{\rm water}$. $U_{\rm water}$ represents the velocity of the jet as it enters the water. All variables in the figure are dimensionless.}
	\label{fig:st}
\end{figure}

Since the piercing jet undergoes rupture during propagation, only the fluid downstream of the rupture point contributes energy to penetration. Typically, the length of this jet segment measures approximately $0.5S_{\rm max}$, where $S_{\rm max}$ denotes the maximum penetration distance. 
Assuming this segment of the jet enters the water at a constant velocity $U_{\rm water}$, and considering the experimental observation that the cross-sectional area of this jet segment remains nearly unchanged, the liquid-bullet's effective mass simplifies to $m = 0.5 \rho AS_{\rm max}$, yielding $K = 1/S_{\rm max}$. We measured $K$ and $U_{\rm water}$ for each case in Figures \ref{fig:16}, \ref{fig:17}, and \ref{fig:18} and the predicted curves from equation (\ref{eq:penetration_depth}) are compared with experimental curves in Figure \ref{fig:st}. The liquid-bullet model predictions agree well with experimental results prior to cavity collapse at the jet tip. After the collapse, the jet's advancing energy dissipates significantly, reaching maximum penetration distance $S_{\rm max}$. Extensive experimental data reveal that collapse times for all piercing jets consistently occur within $0.8 < t < 1.2$. We uniformly adopt $t = 1$ and substitute it into equation (\ref{eq:penetration_depth}) to obtain $S_{\rm max} = U_{\rm water}/\rm ({\it e}-1)$, which represents a general conclusion applicable to most piercing jets. Once the jet velocity $U_{\rm water}$ is determined, we can estimate the maximum penetration distance $S_{\rm max}$ and further obtain the temporal curve of the penetration distance $S$ through equation (\ref{eq:penetration_depth}).

\begin{figure}
	\centerline{\includegraphics[width=0.9\textwidth]{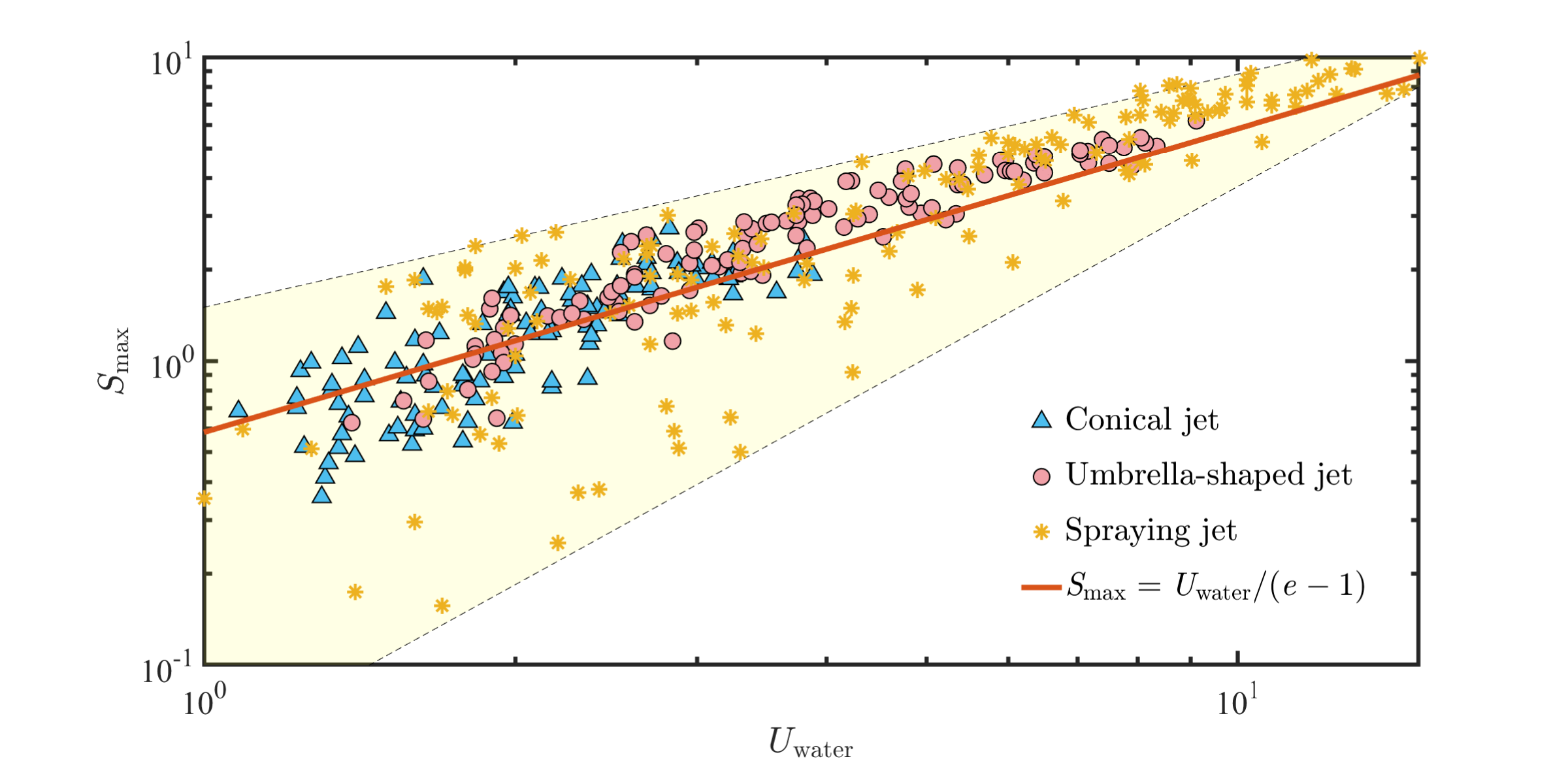}}
	\caption{Variation in the external jet velocity \textit{$U_{\rm water}$} as a function of the maximum penetration distance $S_{\rm max}$. \textit{$U_{\rm water}$} is the velocity of the jet as it enters the water. $S_{\rm max}$ is the distance traveled by the jet in water when the the jet tip cavity collapse. The data collapse onto the red solid line $S_{\rm max} = U_{\rm water}/\rm ({\it e}-1)$, which is derived from equation (\ref{eq:penetration_depth}). The yellow shaded region represents the distribution range of the spraying jet.}
	\label{fig:19}
\end{figure}

We provide a comprehensive insight into the penetration capabilities of the three types of jets in Figure \ref{fig:19}. In general, the penetration capability of a piercing jet is predominantly governed by its external jet velocity $U_{\rm water}$, in agreement with our liquid-bullet model as demonstrated by the solid curves in the figure. The conical and umbrella-shaped jets exhibit a concentrated data distribution. For identical external jet velocities, conical and umbrella-shaped jets achieve comparable maximum penetration distances. However, the umbrella-shaped jet can achieve higher external jet velocities, thereby extending its penetration range. The penetration capability of the spraying jets at low velocity, particularly mist-like spraying jets with pronounced instability, is unstable and slightly lower than that of the conical and umbrella-shaped jets.
This may be due to the excessive measurement of the external jet velocity $U_{\rm water}$ of the mist-like spraying jets. The discontinuous cavity formed by the impact of the fragmented jet tip disturbs the velocity measurement of the upstream continuous jet, leading to the overestimation of the true velocity of the penetration-capable continuous phase.
In contrast, high-velocity spraying jets still adhere to the liquid-bullet model. Their neck breakup acceleration mechanism enables external velocities exceeding those achievable by umbrella-shaped jets. Regarding the upper limit of jet penetration capability, spraying jets outperform the other types, making them highly suitable for applications requiring long-distance penetration and efficient material transport. For scenarios demanding shorter penetration distances, umbrella-shaped jets, owing to their superior stability and controllability, represent a more optimal choice. The methods for controlling initial parameters to achieve umbrella-shaped jets or optimize the penetration performance of spraying jets will be discussed in the following section.

\begin{figure}
	\centering
	\begin{subfigure}[htbp]{0.49\textwidth}
		\begin{overpic}[width=\textwidth,trim=20pt 0 60pt 0, clip]{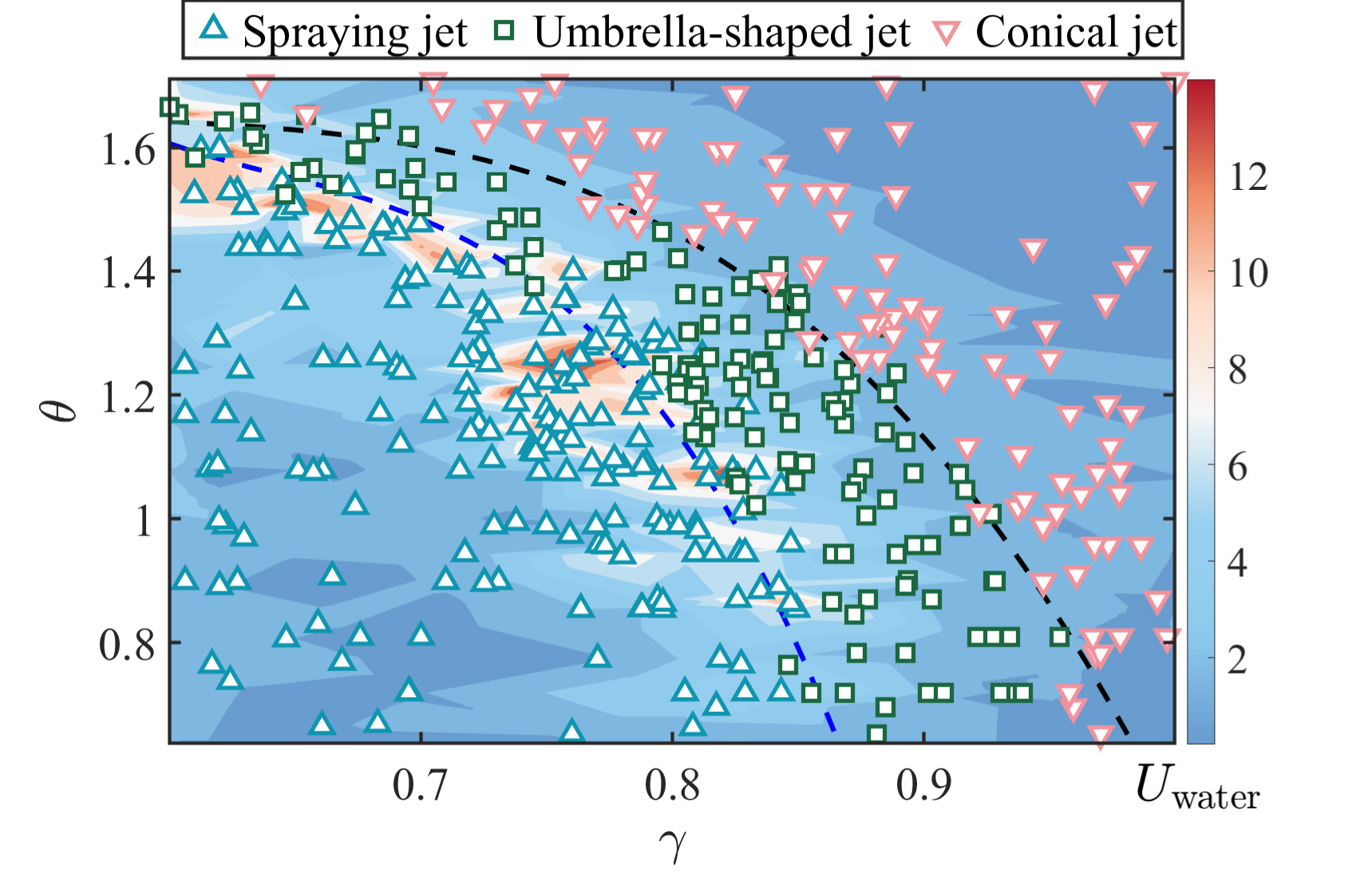} 
			\put(5,67){\large \textrm{(\it a\rm)}} 
		\end{overpic}
	\end{subfigure}
	\vspace{-0.02\textwidth} 
	\begin{subfigure}[htbp]{0.49\textwidth}
		\begin{overpic}[width=\textwidth,trim=20pt 0 60pt 0, clip]{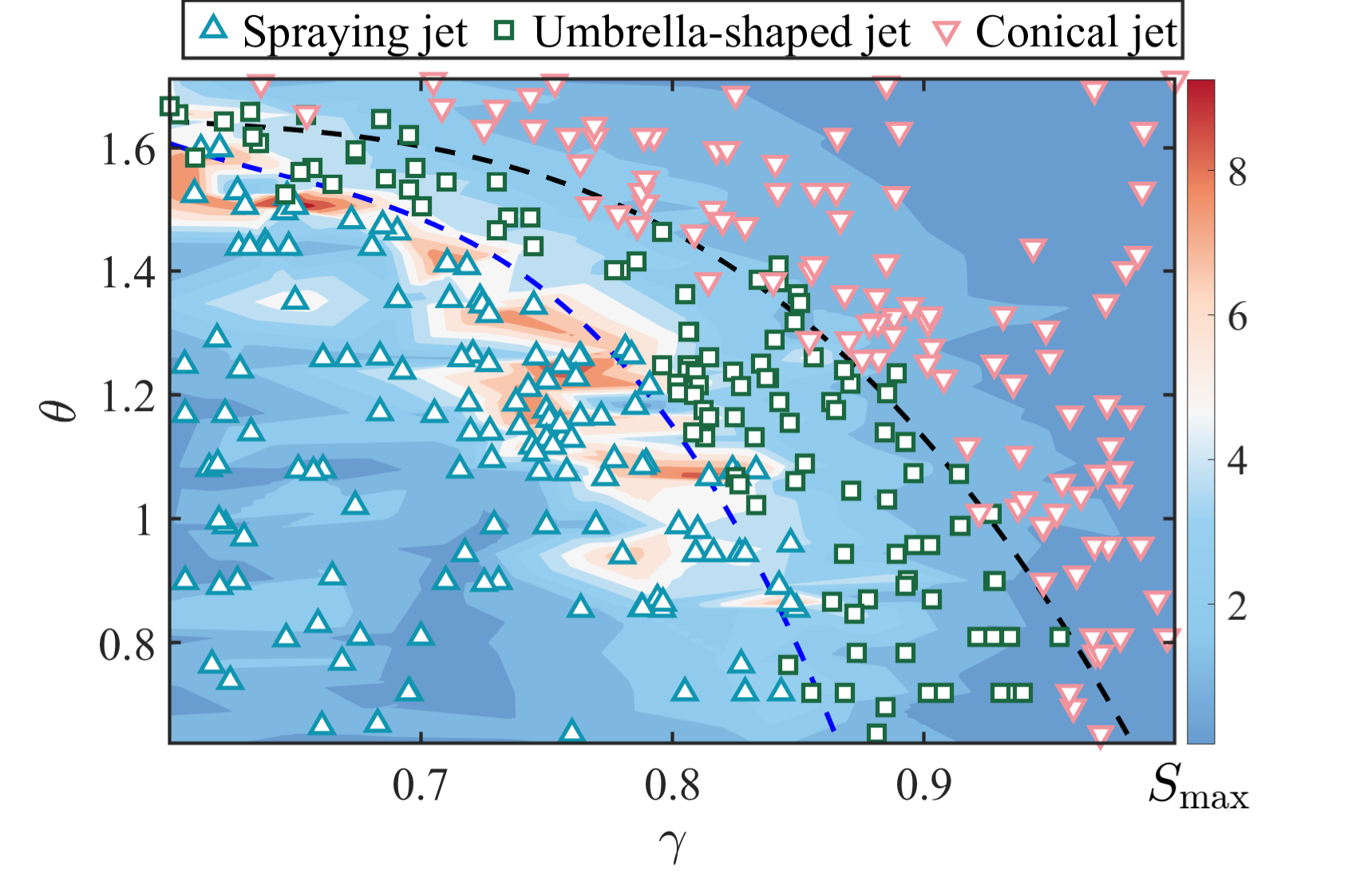} 
			\put(5,67){\large \textrm{(\it b\rm)}} 
		\end{overpic}
		\vspace{-0.05\textwidth} 
	\end{subfigure}
	\caption{Phase diagram of the external jet velocity \textit{$U_{\rm water}$} ($a$) and the maximum penetration distance $S_{\rm max}$ ($b$) as a function of $\gamma$ and $\theta$, with background color interpolated from experimental data. The dashed lines represent the fitted boundaries between different jet regimes: the black dashed line separates conical jets and umbrella-shaped jets, while the blue dashed line distinguishes umbrella-shaped jets from spraying jets.}
	\label{fig:20}
\end{figure}

\subsection{Maps of jet regime, velocity and penetration distance in $\theta-\gamma$ parameter space}\label{6.2}
This section examines the effect of initial parameters $\theta$ and $\gamma$ of the tandem bubbles system on the different regimes of the piercing jet. Approximately 300 experimental data points are selected and interpolated to obtain the phase diagrams, where distinct symbols differentiate the three piercing jet regimes across the investigated parameter space ($0.7 < \theta < 1.7$ and $0.6< \gamma < 1$).
Particular emphasis is placed on the external jet velocity $U_{\rm water}$ and maximum penetration distance $S_{\rm max}$ of the piercing jet after penetrating the bubble surface.

Figures \ref{fig:20}($a$) and ($b$) present phase diagrams of three jet regimes according to $\theta$ and $\gamma$. Our experimental data primarily focus on parameter space generating high velocity spraying jets and umbrella-shaped jets, given their critical relevance to needle-free injection applications. As the relative initiation time difference $\theta$ increases from 0, the interaction between bubble 1 and bubble 2 transitions from attraction to repulsion, causing the jet direction to reverse from pointing toward bubble 1 to penetrating through bubble 2. When $\theta$ exceeds approximately 2, bubble 2 becomes hydrodynamically decoupled from bubble 1's primary oscillation. The dynamics of bubble 2 are governed solely by pressure waves from bubble 1's collapse and subsequent re-expansion. Therefore, this decoupled regime falls outside the scope of our investigation. The background color in each figure represents the magnitude of $U_{\rm water}$ and $S_{\rm max}$, with different symbols used to distinguish three piercing jet regimes. The boundaries between these regimes are denoted by dashed lines: the black dashed line separates conical jets from umbrella-shaped jets, while the blue dashed line distinguishes umbrella-shaped jets from spraying jets.

Figure \ref{fig:20}(\textit{a}) demonstrates that the jet regime is governed by the interplay of the parameters $\theta$ and $\gamma$. Larger $\theta$ and $\gamma$, which correspond to greater relative initiation distances and time differences, facilitate the formation of conical jets, while smaller values predominantly result in spraying jets. The stable umbrella-shaped jets are confined to a narrow intermediate region between these two regimes. The upper right and lower left regions of Figure \ref{fig:20}(\textit{a}) correspond to regimes of minimal and maximal bubble-bubble coupling, respectively. These regions exhibit lower jet velocities, characteristic of conical jets and mist-like spraying jets. The spraying jets and umbrella-shaped jets with high velocity, which are suitable for achieving underwater directional energy transfer, are widely distributed in the region with moderate values of $\theta$ and $\gamma$. The high-speed jet region of our phase diagram reproduces the findings reported by \citet{han2015dynamics} and \citet{fan2024amplification}. With substantially more experimental data points, our phase diagram achieves higher resolution than prior works. We find that the high-speed jet region in the phase diagram exhibits a band-like distribution along the left side of the boundary between umbrella-shaped jets and spraying jets. Parameter regions generating internal velocities exceeding 1200 m/s are mapped, while their corresponding external velocity variations are simultaneously quantified. This enhanced resolution enables higher-precision in controlling piercing jets.

Since the external jet velocity $U_{\rm water}$ cannot fully characterize the penetration capability of the piercing jets, we further investigate the distribution of their maximum penetration distance $S_{\rm max}$ within the $\theta-\gamma$ parameter space. Its distribution pattern within the $\theta-\gamma$ parameter space resembles that of the external velocity, as shown in Figure \ref{fig:20}(b). Spraying jets exhibit optimal penetration performance for long-range applications within the red-highlighted parameter region, which corresponds to the needle-like spraying jet. Meanwhile, the umbrella-shaped jets adjacent to the red-highlighted region represent superior short-range penetration capability. These findings establish a comprehensive control framework for utilizing tandem-bubble jets as a viable alternative to conventional needle-based injection systems.

\section{Summary and conclusions}\label{8}

This study reveals the dynamics of piercing jets produced by two anti-phase, centimetre-sized tandem cavitation bubbles. By controlling the relative initiation distance $\gamma$ and relative initiation time difference $\theta$ between the two bubbles, we have experimentally identified three distinct regimes of piercing jets, namely, conical jets, umbrella-shaped jets, and spraying jets. These experimental observations are complemented by numerical simulations utilizing the Volume of Fluid (VoF) method and Boundary Integral Method (BIM), which elucidate the underlying mechanisms governing the formation of each type of piercing jets. The key findings are summarized as follows.

For an anti-phase bubble pair, the over-expanded bubble 1 (radius beyond equilibrium) attracts bubble 2 and draws its near side into a high-curvature, elongated tip. This tip collapses first and evolves into a slender piercing jet whose dynamics are set by the spatiotemporal mismatch between the final collapse of bubble 1 and the self-contraction of the tip on bubble 2. If bubble 1 collapses before the tip can contract under its own curvature, the impinging pressure waves drive the inward flow into a conical jet. The jet velocity scales linearly with the impulse of the pressure wave, indicating that surface curvature initiates the jet, while the pressure wave governs its subsequent growth. The formation of a conical jet is therefore controlled by the interplay between local curvature and the pressure wave. When the tip contracts before bubble 1 collapses, an umbrella-shaped jet emerges. The converging flow accelerates the upstream fluid beyond the velocity of the advancing tip. As this faster fluid overtakes the tip, it sweeps the leading-edge liquid sheet outward into a thin radial film. Should bubble 1 then collapse adjacent to the jet base, the emitted pressure waves can induce a second umbrella-like sheet following the jet tip.

Spraying jets originate from the neck breakup of bubble 2's elongated tip, triggered by the high-pressure waves emitted when bubble 1 collapses. These jets can exceed 1200 m/s and initially exhibit a compact, needle-like morphology. Although the inherent instability of the neck breakup may fragment and tilt the tip, the continuous jet remains highly stable thanks to support from the jet base, exhibiting strong axisymmetry and sustained speed. At sufficiently small relative initiation distances $\gamma$, the needle-like jets are no longer accelerated by the localized high-pressure stagnation point produced by neck breakup. Instead, the elongated tip of bubble 2 coalesces with bubble 1, entraining a cloud of dispersed micro-bubbles during neck contraction. The subsequent collapse of these micro-bubbles generates intermittent high-pressure fluctuations that ultimately produce a chaotic, mist-like spraying jet, markedly different from the initial needle-like structure.

After penetrating the bubble interface, the piercing jet trails an elongated gas cavity behind its tip. The maximum cavity length serves as a robust quantitative indicator of jet penetration capability. Our analysis reveals distinct penetration traits among the different jet morphologies. The needle-like spraying jet exhibits the highest performance, with cavity lengths reaching more than ten times the bubble's maximum radius. The umbrella-shaped jet offers optimal short-range penetration, thanks to its superior flow stability and directional controllability. In contrast, the conical and mist-like spraying jets suffer from limited penetration, owing to their lower velocities and rapid energy dissipation during penetration. Remarkably, the penetration process and maximum distance of all piercing-jet regimes are well predicted by our simplified liquid-bullet model. Through extensive experiments we systematically mapped the formation conditions of the three jet regimes in the $\gamma$-$\theta$ parameter space and quantified their maximum velocities and penetration distances in water. These results provide a comprehensive understanding of jet formation and evolution in anti-phase bubble systems and establish a foundational framework for applications that demand precise control of jet behavior, such as needle-free injection.

\section*{Acknowledgements}
\label{S:8}
We are thankful for the insightful discussion and the help from Q.X. Wang.

\section*{Funding}
\label{S:9}
This work is supported by the National Natural Science Foundation of China (nos. 52525102, 12372239, 12572284), the Key R$\&$D Program Project of Heilongjiang Province (JD24A002), the National Key Laboratory of Ship Structural Safety (Naklas2024ZZ004-J), and the Xplore Prize.

\section*{Declaration of interests}
\label{S:10}
The authors report no conflict of interest.

\section*{Author ORCIDs}
\label{S:11}
Shuai Yan https://orcid.org/0009-0005-3869-7231; A-Man Zhang https://orcid.org/0000-0003-1299-3049; Tianyuan Zhang https://orcid.org/0000-0001-7238-2608; Pu Cui https://orcid.org/0000-0001-8179-1366; Rui Han https://orcid.org/0000-0003-3699-5954; Shuai Li https://orcid.org/0000-0002-3043-5617.

\section*{Appendix}
\label{S:12}
\begin{figure} 
	\centerline{\includegraphics[trim={0pt 0 0pt 300pt}, 
		clip, width=0.8\textwidth]{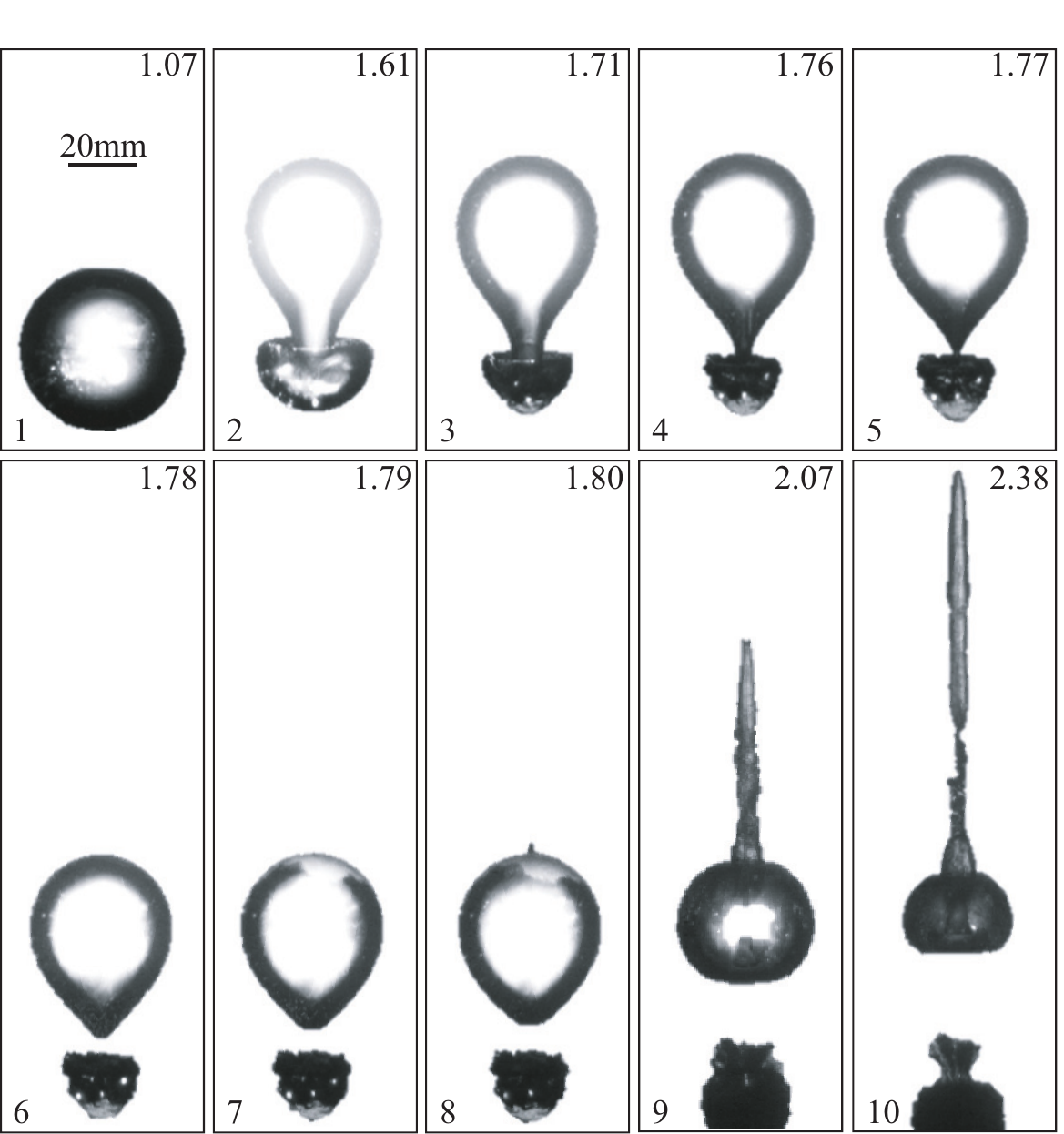}}
	\caption{Dynamics of spraying jet from anti-phase bubble pair, corresponding to the case exhibiting the highest jet velocity in Figure \ref{fig:6}($a$). The time scale $ R_{\rm max} \sqrt{\rho / P_{\rm \infty}}$ is 2.25ms. The initial parameters are $\gamma=0.78$, $\theta=1.07$.}
	\label{fig:appendix}
\end{figure}

This section presents the image sequences corresponding to the case exhibiting the highest jet velocity in Figure \ref{fig:6}($a$). As the Figure \ref{fig:appendix} shown, from $t \approx 1.77$ (frame 5) to $t \approx 1.79$ (frame 7), the elongated tip of bubble 2 undergoes neck breakup, after which the jet tip impacts the upper surface of bubble 2. Over this 0.02 interval, the jet travels a total distance of 2.49, yielding a minimum estimated velocity of 123 (corresponding to 1230 m/s).

\bibliographystyle{jfm}
\bibliography{jfm1}

\expandafter\ifx\csname natexlab\endcsname\relax
\def\natexlab#1{#1}\fi
\expandafter\ifx\csname selectlanguage\endcsname\relax
\def\selectlanguage#1{\relax}\fi

\end{document}